\documentclass[]{pasj02}
\usepackage{bm}
\usepackage{url}
\usepackage{pifont}
\usepackage{ulem}
\usepackage{physics}

\jyear{2024}
\Received{}
\Accepted{}

\newcommand{\Msun}{M_{\odot}}
\newcommand{\Zsun}{Z_{\odot}}
\newcommand{\na}{New Astron.}          
\newcommand{\nar}{New Astron. Rev.}    
\newcommand{\rmxaa}{Rev. Mex. Astron. Astrofis.} 
\newcommand{\pasa}{Publ. Astron. Soc. Australia} 

\begin{document}

\title{Bursty star formation, chemical enrichment, and star cluster formation in
numerical analogues of GN-z11}

\author{
Takayuki \textsc{R. Saitoh},\altaffilmark{1,2}\altemailmark\orcid{0000-0001-8226-4592} \email{saitoh@people.kobe-u.ac.jp}
Yutaka \textsc{Hirai},\altaffilmark{3}\orcid{0000-0002-5661-033X}
Michiko \textsc{S. Fujii},\altaffilmark{4}\orcid{0000-0002-6465-2978}
 and
Yuki \textsc{Isobe}\altaffilmark{5,6,7}$^{,\dag}$\orcid{0000-0001-7730-8634}
}
\altaffiltext{1}{Department of Planetology, Graduate School of Science, Kobe University 1-1 Rokkodai, Nada-ku, Kobe, Hyogo 657-8501, Japan}
\altaffiltext{2}{Center for Planetary Science (CPS), Graduate School of Science, Kobe University 1-1 Rokkodai, Nada-ku, Kobe, Hyogo 657-8501, Japan}
\altaffiltext{3}{Department of Community Service and Science, Tohoku University of Community Service and Science, 3-5-1 Iimoriyama, Sakata, Yamagata 998-8580, Japan}
\altaffiltext{4}{Department of Astronomy, Graduate School of Science, The University of Tokyo, 7-3-1 Hongo, Bunkyo-ku, Tokyo 113-0033, Japan}
\altaffiltext{5}{Kavli Institute for Cosmology, University of Cambridge, Madingley Road, Cambridge, CB3 0HA, UK}
\altaffiltext{6}{Cavendish Laboratory, University of Cambridge, 19 JJ Thomson Avenue, Cambridge, CB3 0HE, UK}
\altaffiltext{7}{Waseda Research Institute for Science and Engineering, Faculty of Science and Engineering, Waseda University, 3-4-1, Okubo, Shinjuku, Tokyo 169-8555, Japan}

\footnotetext[$\dag$]{JSPS fellow.}


\maketitle

\begin{abstract}
The James Webb Space Telescope reveals anomalous nitrogen enrichment (high N/O
ratios) in compact, star-forming galaxies, such as GN-z11 at $z\sim10$.  The
origin of this chemical signature provides an insight into the early star
and galaxy formation processes, yet remains unclear.  We performed
high-resolution cosmological zoom-in simulations of massive galaxies at high
redshift ($z\sim10$) in rare density peaks, incorporating various chemical
evolution channels including stellar winds, core-collapse, Type Ia
supernovae, and asymptotic giant branch stars.  Our simulations reproduce
several key features of high-redshift galaxies: (1) stars form with high
efficiencies ($>0.1$) at the center of rare peak halos, creating very
compact galaxies similar to GN-z11; (2) high N/O ratios emerge during the
first 10-20 Myr of intense starburst, before being diluted by CCSNe; (3)
multiple star clusters form in and around the galaxy with high efficiency
($\sim20\%$), some of which exhibit high N/O ratios and sodium-oxygen
anti-correlations similar to those observed in local globular clusters.
Although our simulations can reproduce the high log(N/O) values (up to -0.61,
exceeding the solar value by 0.25 dex), they remain below the observational
lower limits of GN-z11, indicating room for improvement through additional
chemical evolution channels, such as supermassive stars.
\end{abstract}


\section{Introduction} \label{sec:intro}

The James Webb Space Telescope (JWST) provides a number of high-redshift ($z$)
galaxies (e.g., \cite{Pontoppidan+2022}). Some of them are spectroscopically
confirmed as galaxies at $z> 10$ (e.g., \cite{Robertson+2023,
Curtis-Lake+2023,ArrabalHaro+2023Nature, Robertson+2024JOF, Finkelstein+2024,
Harikane+2025,Hainline+2024,Finkelstein+2025}).  These observations have
activated discussions about the early phase of galaxy formation.
\citet{Adamo+2025ConferenceSummary} provide a recent overview of JWST's impact on
our understanding of high-$z$ galaxy formation.  Among them, the detections of
the nebular emissions, such as helium (He), carbon (C), nitrogen (N), oxygen (O),
neon (Ne), sulfur (S), give us detailed insights into the formation process of
high-$z$ galaxies. These observations are important clues to understanding the
formation and evolution of galaxies, particularly in their early phases.

GN-z11 is one of the high-$z$ galaxies in the range of $z > 10$. GN-z11 was
originally discovered by the Hubble Space Telescope \citep{Oesch+2016} and its
redshift was precisely confirmed with JWST \citep{Bunker+2023,Tacchella+2023}.
The spectroscopic observations reveal that this galaxy has unique chemical
compositions which might come from the specific environment of the early phase
of galaxy formation. \citet{Cameron+2023} report that GN-z11 has a high log(N/O)
value, more than four times of the solar value. \citet{Senchyna+2024} show that a
high electron number density is required to achieve a high log(N/O) value. This high
log(N/O) value with high electron number density may be a unique signature of
the ISM around dense star-cluster formation regions.

Including GN-z11, twelve nitrogen-enhanced galaxies have been found at $z\ge6$:
CEERS 397 \citep{Sanders+2024, Stiavelli+2025}, 
RXCJ2248-4431 \citep{Topping+2024}, 
GLASS-150008 \citep{Isobe+2023}, 
A1703-zd6 \citep{Topping+2025}, 
GN-z8-LAE \citep{Navarro-Carrera+2025}, 
CEERS-1019 \citep{Isobe+2023,Marques-Chaves+2024}, 
GN-z9p4 \citep{Schaerer+2024}, 
JADES-GS-z9-0 \citep{Curti+2025JADES-GS-z9-0},
GHZ9 \citep{Napolitano+2025,Atek+2023,Castellano+2023},
GN-z11 \citep{Cameron+2023, Bunker+2023},
GHZ2/GLASS-z12 \citep{Castellano+2024, Zavala+2025Nat},
MoM-z14 \citep{Naidu+2025}.
Their properties are summarized in table \ref{tab:Nenhanced}. These galaxies
share common characteristics, although there are some outliers: they are massive
($\sim10^{8-9}~\Msun$), young ($1-30~\mathrm{Myr}$), compact
($10-100~\mathrm{pc}$), and low-metallicity ($\sim 0.1~\Zsun$) objects having
high star formation rates (SFRs;$5-160~\Msun~\mathrm{yr}^{-1}$).  While their
log(N/O) values are elevated, their log(C/O) values remain at normal levels. The
fraction of nitrogen-enhanced galaxies is so far estimated as $\sim 1-2\%$ in
the high-$z$ galaxy samples of $\sim 500$ \citep{Schaerer+2024}. Readers can see
a similar table with earlier samples in \citet{Rizzuti+2024}.
\citet{Harikane+2025} find that bright galaxies at $z\gtrsim10$ can be
classified into two groups: extended and compact galaxies.  The compact group
shows nitrogen enhancements.

\begin{table*}
\tbl{N-enhanced galaxies at $z\ge6$ \label{tab:Nenhanced}}{%
\begin{tabular}{lcccccccc}
\hline
Name &
$z$ &
$\log(M_\mathrm{star}/\Msun)$ &
SFR$({\Msun~\mathrm{yr^{-1}}})$ &
Age(Myr) &
$R_{\mathrm{eff}}$(pc) &
log(O/H)+12 &
log(N/O) &
log(C/O) \\
\hline
CEERS 397 & 6.000 & 9.63 & 42.7 & -- & $420\pm20$ & $7.96^{+0.10}_{-0.08}$ & $-0.67^{+0.23}_{-0.17}$ & -- \\
RXCJ2248-4431 & 6.1057 & ${8.05^{+0.17}_{-0.15}}$ & $63^{+37}_{-23}$ & $1.8^{+0.7}_{-0.4}$ &
$<22$ & $7.43^{+0.17}_{-0.09}$ & $-0.39^{+0.10}_{-0.08}$ & $-0.83^{+0.11}_{-0.10}$ \\
GLASS-150008  & 6.23 & ${8.39^{+0.35}_{-0.19}}$ & $5.1^{+6.2}_{-1.1}$& $126^{+375}_{-70}$ & $400$\footnotemark[$*$] & $7.65^{+0.14}_{-0.08}$ &
$-0.40^{+0.05}_{-0.07}$ & $-1.08^{+0.06}_{-0.14}$\\
A1703-zd6 & 7.0435 & $7.70\pm0.24$ & $30.9^{+29.4}_{-13.9}$ & $1.6^{+0.5}_{-0.4}$ & $<120$ & $7.47\pm0.19$ & $-0.6\pm0.3$ & $-0.74\pm0.18$\\
GN-z8-LAE & 8.279 & ${7.66\pm 0.02}$ & $11.3\pm 7.8$ & $2-6$ & $143\pm15$ & $7.85\pm0.17$ &
$-0.44\pm0.36$ & $-0.69\pm0.21$\\
CEERS-1019  & 8.6791 & ${9.31\pm0.15}$ & $161\pm23$ & $14.4\pm7.2$ & $100-145$ & $8.23-8.50$ &
$>-0.01$ & $<-0.36$\\
GN-z9p4  & 9.380 & ${8.7\pm0.2}$ & $64\pm14$ & -- & $118\pm16$& $7.37\pm0.15$ & $-0.58\pm0.2$  & $<-1.18$  \\
JADES-GS-z9-0  & 9.4327 & ${8.17^{+0.02}_{-0.04}}$ & $4.34^{+0.1}_{-0.08}$ & ${2.80^{+0.06}_{-0.03}}$\footnotemark[$\dagger$] & $110\pm9$& $7.49\pm0.11$ & $-0.93\pm0.24$  & $-0.90\pm0.12$ \\
GHZ9  & 10.145 & ${8.69^{+0.24}_{-0.46}}$ & $0.56^{+0.23}_{-0.29}/14.4^{+15.0}_{-7.3}$ & -- & $<106$ & $6.7-7.7$  & $ -0.08\sim0.12$ & $-0.96\sim-0.45$ \\
GN-z11  & 10.60 & ${9.1^{+0.3}_{-0.4}}$ & $21^{+22}_{-10}$ & $24^{+20}_{-10}$ & $64\pm20$& $7.91\pm0.07$ & $>-0.25$ & $>-0.78$ \\
GHZ2/GLASS-z12  & 12.34 & ${9.05^{+0.10}_{-0.25}}$ & $5.2^{+1.1}_{-0.6}$ & $28^{+10}_{-14}$\footnotemark[$\ddagger$] & $105\pm9$ & $7.26^{+0.27}_{-0.24}$ & $-0.29\sim-0.2$  & $-0.94\sim-0.53$  \\
MoM-z14 & $14.44$  & $8.1^{+0.3}_{-0.2}$ & $13.0^{+3.7}_{-3.5}$ & $4.0^{+10.0}_{-1.4}$ & $74^{+15}_{-12}$& $7.33^{+0.65}_{-0.56}$ & $-0.57^{+0.28}_{-0.45}$ & $-0.91^{+0.39}_{-0.22}$ \\
\hline
\end{tabular}}
\begin{tabnote}
References are as follows: CEERS 397\citep{Mascia+2024, Sanders+2024,
Stiavelli+2025}, RXCJ2248-4431\citep{Topping+2024},
GLASS-150008\citep{Isobe+2023, Jones+2023, Mascia+2024},
A1703-zd6\citep{Topping+2025}, GN-z8-LAE\citep{Navarro-Carrera+2025},
CEERS-1019\citep{Isobe+2023,Marques-Chaves+2024},
GN-z9p4\citep{Schaerer+2024}, JADES-GS-z9-0\citep{Curti+2025JADES-GS-z9-0},
GHZ9 \citep{Napolitano+2025,Atek+2023,Castellano+2023},
GN-z11\citep{Cameron+2023,Tacchella+2023,Alvarez-Marquez+2025},
GHZ2/GLASS-z12\citep{Castellano+2024}, MoM-z14 \citep{Naidu+2025}. \\
\footnotemark[$*$] Evaluated under the minimum scale of $\sim140~\mathrm{pc}$ at $z=6$ \citep{Mascia+2024}. \\
\footnotemark[$\dagger$] Light-weighted age evaluated by {\tt BEAGLE}. \\
\footnotemark[$\ddagger$] Mass-weighted age.
\end{tabnote}
\end{table*}

The discovery of this uniqueness has triggered active discussion about the
origin of this chemical abundance. Several pollution mechanisms have been
proposed: (1) stellar wind from massive, fast-rotating stars [Wolf-Rayet (WR)
stars] (e.g., \cite{Cameron+2023,BekkiTsujimoto2023,KobayashiFerrara2024,
Watanabe+2024, FukushimaYajima2024}), (2) stellar wind from very/supermassive
stars (e.g., \cite{Charbonnel+2023, NageleUmeda2023, Vink2023, Nandal+2024}),
(3) asymptotic giant branch (AGB) stars (e.g., \cite{Cameron+2023,
D'Antona+2023}), and (4) tidal disruption events (e.g., \cite{Cameron+2023}).
There is no clear consensus on which mechanism is responsible.  Most prior
studies employ one-zone or semi-analytical models that cannot self-consistently
treat the complex interplay among gas accretion, star formation, feedback, and
chemical evolution. This limitation makes it difficult to quantify the dominant
mechanism under the realistic galaxy formation conditions.

The high log(N/O) value is also found in local globular clusters (GCs)
(e.g., \cite{Carretta+2005, BastianLardo2018, MiloneMarino2022}). This implies
that the high-$z$ galaxy formation is closely linked to GC formation
\citep{BelokurovKravtsov2023, D'Antona+2023}.  Through gravitational lensing
observations, \citet{Adamo+2024} discovered high surface density star clusters
densely populated within a region of $70~\mathrm{pc}$ at $z = 10.2$. These
clusters are very young and have very low metallicities ($\sim5\%$ of the solar
metallicity).  Similarly, \citet{Hsiao+2023} observed highly lensed two star
clusters at $z\sim10$ and \citet{Mowla+2024} discovered ten dense and compact
star clusters at $z=8.296$, indicating that star cluster formation was common
during this epoch. \citet{Senchyna+2024} argued that, based on comparisons with
nearby galaxies containing high concentrations of WR stars, the nitrogen
enrichment may indicate a signature of the dense clustered star formation in
GN-z11. Supporting this connection, \citet{Rivera-Thorsen+2024} detected WR
stars in a strongly gravitationally lensed galaxy, the Sunburst Arc, at $z =
2.37$. This galaxy contains a young ($\sim 4~\mathrm{Myr}$), massive
($9\times10^6~\Msun$) star cluster whose spectrum shows a significant
enhancement in the log(N/O) value.

Galaxy formation and star cluster formation are fundamentally linked during the
early stages of galaxy formation.  Chemical evolution also serves as an
important guideline to arrange their formation processes.  Theoretically,
\citet{Dekel+2023} has constructed a galaxy formation model with highly efficient
star formation at $z\sim10$. Their study compared free-fall time with feedback
ignition time, and pointed out that efficient star formation, starburst, occurs
in galactic centers under conditions where $n_\mathrm{th} > 3\times10^3~\mathrm
{cm}^{-3}$, when the free-fall time becomes shorter than the characteristic
feedback timescale of 1 Myr (feedback-free starburst).  Furthermore, star
cluster formation is shown to accompany this process based on the Jeans mass at
the galactic center. Since galaxy formation involves complicated non-linear
processes, including chemical evolution, it is important to elucidate formation
processes through simulations while comparing them with theoretical models.  Our
simulations test whether this feedback-free starburst scenario is realized in
the galaxy formation process, and can naturally explain the observational
properties of N-enhanced galaxies, in their compact sizes, high SFRs, and
chemical signatures.

In this study, we present high-resolution cosmological simulations within a
large volume ($\sim 100~{\rm Mpc}^3$) and fine spatial resolution ($\sim
5~\mathrm{pc}$) that, to our knowledge, are among the first to simultaneously
address three fundamental questions: (1) How do compact galaxies like GN-z11
emerge in the early universe around $z\sim10$?  (2) What process is responsible
for the observed high log(N/O) values?  and (3) How are galaxy formation and
star cluster formation connected?  Our simulations demonstrate that stellar
winds from fast rotating massive stars can naturally explain high log(N/O)
values during starburst phase at the galactic center, which has a compact
structure consistent with the feedback-free starburst scenario.

The structure of this article is the following: in section~\ref{sec:method}, we
explain our model of numerical simulations. Section~\ref{sec:results} describes
the numerical results of the fiducial model. Model comparisons are presented in
Section~\ref{sec:comparisons}.  Summary and discussion are provided in
Section~\ref{sec:summary}.

\section{Methods}\label{sec:method}

\subsection{Initial condition}

According to \citet{Scholtz+2024}, the expected dark matter halo mass of GN-z11
is $\sim 3 \times 10^{10}~\Msun$. This is comparable with the expected mass of
the Coma-like cluster at that epoch via the abundance matching method.
Therefore, as a first attempt, we consider GN-z11 to be the progenitor of a
Coma-like cluster.  We generated an initial condition of such a rich cluster
using a zoom-in technique.

The initial conditions for our zoom-in cosmological simulations were generated
using {\tt MUSIC} \citep{HahnAbel2011}. We adopted the cosmological parameters
from Planck 2018 \citep{Planck2018}: $H_0 = 67.32$, $\Omega_{\mathrm{M}} =
0.3158$, $\Omega_{\mathrm{\Lambda}} = 0.6842$, $\Omega_{\mathrm{b}} = 0.04938$,
and $\sigma_8 = 0.812$. The simulation box was set to $100~{\mathrm{Mpc}}~h^{-1}
$ on each side.  The initial redshift was set to 100.

First, we performed a dark matter only simulation with $256^3$ particles
(level=8) under a periodic boundary condition using {\tt Gadget-2}
\citep{Springel2005}. We then used the AMIGA Halo Finder (AHF:
\cite{Knollmann+2009}) to make halo catalogs and generated the merger trees to
investigate the detailed mass growth histories. From this analysis, we
identified the largest halo of which mass is $\sim 10^{10}~\Msun$ at $z\sim10$.

We zoomed in the halo using the snapshot of $z \sim 8$. We adopted an ellipsoid
region that included all particles within four times the virial radius of the
target halo. To prevent contamination from the boundary particles, we added an
expansion factor of 1.05 to the ellipsoidal region \citep{Griffen+2016}. The
finest grid was set to level 14.

Table \ref{tab:number} summarizes the initial number of particles and
gravitational softening lengths.  The softening lengths are proportional to the
scale factor for $z > 9$ and fixed at the listed values for $z\leq9$.

Based on our target galaxy's number density of $\sim 10^{-6}
~{\mathrm{Number~Mpc^{-3}}}$, the UV absolute magnitude expected from
the UV luminosity function at $z\sim10$ is $M_{\mathrm{UV}} = -22\sim-21$
\citep{Harikane+2025}. This is comparable to that of GN-z11, $M_{\rm UV} =
-22.1\pm0.2$ \citep{Oesch+2016}.

\begin{table*}
\tbl{Initial number of particles and mass/spatial resolutions\label{tab:number}}{
\begin{tabular}{ccccccccc}
\hline
Level & $N_\mathrm{Baryon}$ & $N_\mathrm{DM}$ & $N_\mathrm{Boundary}$ & $N_\mathrm{Total}$ &
$m_\mathrm{baryon}$ & $m_\mathrm{DM}$ & $\epsilon_\mathrm{baryon}$ & $\epsilon_\mathrm{DM}$\footnotemark[a] \\
\hline
14&83892632&83892632&21948083 & 189733347 & $4628~\Msun$  &$24972~\Msun$ & $5.7~\mathrm{pc}$ & $12.6~\mathrm{pc}$ \\
\hline
\end{tabular}}
\begin{tabnote}
\footnotemark[a] The quantities shown in this column are the minimum
gravitational softening lengths. Boundary particles have larger softening
lengths corresponding to their masses.
\end{tabnote}
\end{table*}

\subsection{The code}
We use our original parallel $N$-body/smoothed particle hydrodynamics (SPH) code,
{\tt ASURA} \citep{Saitoh+2008,Saitoh+2009} for our simulations. The code solves
gravitational interactions using the tree method \citep{BarnesHut1986}. The
parallelization strategy follows \citet{Makino2004}. We adopt the symmetrized
Plummer potential to evaluate gravity \citep{SaitohMakino2012}. The opening
angle is set to 0.5 and only the monopole term is taken into account. We use
{\tt Phantom-GRAPE} to accelerate gravity computations \citep{Tanikawa+2013}.

For hydrodynamical interactions, we adopt density independent SPH (DISPH)
\citep{SaitohMakino2013}, an improved version of the original SPH
\citep{Lucy1977,GingoldMonaghan1977} that better handles fluid instabilities.
Here, we use the Wendland C4 kernel for convolutions \citep{DehnenAly2012} with
a neighbor number of $128\pm8$. To handle shocks, we implement an artificial
viscosity term following \citet{Monaghan1997} combining it with a variable
viscosity parameter method proposed by \citet{Rosswog2009} and the Balsara switch
\citep{Balsara1995}.

We employ the hierarchical and block timestep methods \citep{McMillan1986,
Makino1991IndividualTimeStep}. To reduce errors induced by large timestep
differences in hydrodynamic interaction regions, we limit the timestep
difference between gas particles to less than four times \citep{SaitohMakino2009}
. To accelerate computation, we adopt the FAST method \citep{SaitohMakino2010}.
This method reduces the computational cost in supernovae-induced hot regions by
assigning different timesteps for gravitational and hydrodynamical interactions.

We use {\tt Grackle} \citep{Smith+2017} to solve the radiative cooling and
heating.  We track twelve primordial species (H, H$^+$, He, He$^+$, He$^{++}$,
$e$, H$_2$, H$^-$, H$_2^{+}$, D, D$^{+}$, HD) and employ the pre-computed {\tt
Cloudy} tables (Ver. 13.05; \cite{Ferland+2013}) for metal cooling.  Our
simulations include both the ultraviolet background \citep{HaardtMadau2012} and
the redshift evolution of the mean Lyman-Werner background (LWB) as evaluated by
\citet{Incatasciato+2023}.{\footnote{We used the equation (9) from their paper
and combined it with {\tt Grackle}. The strength of the ``mean'' LWB was
evaluated by using a post-processing method applied to the FiBY project
simulations \citep{Johnson+2013}.}} We introduce a normalization parameter
$f_\mathrm{LWB}$ to scale the strength of LWB, where $f_\mathrm{LWB} = 1.0$ is the
default value. These heating effects are time-dependent, but spatially uniform.

When a gas particle satisfies star formation criteria, it spawns a star particle
which has the same metallicity of the parent gas particle in a probabilistic
manner following the Schmidt law. We adopt star formation criteria similar to
those used in \citet{Saitoh+2008, Saitoh+2009}: (1) high-density ($n_{\rm H} >
1000~\mathrm{cm}^{-3}$), (2) low-temperature ($T <100~\mathrm{K}$), and (3)
converging flow ($\div \bm{v} < 0$).  The star formation efficiency,
$C_{*}$ is set to 0.5.  The typical mass of star particles is one-third of the
initial gas particle mass. If a mass of a gas particle becomes less than
one-third of the initial gas particle mass, it converts directly into a star
particle instead of spawning a new one. As an exceptional case, when the gas
density exceeds $10^5~\mathrm{cm}^{-3}$ (i.e., 100 times the threshold density),
the gas particle is immediately converted to a star particle regardless of other
criteria.

We assume a simple stellar population model for star particles.  For a
metal-poor gas ($Z<10^{-5} \Zsun$), we adopt the Susa initial mass function (IMF)
\citep{Susa2013} from $0.7$ to $300~\Msun$ and otherwise adopt the
Chabrier IMF \citep{Chabrier2003} from $0.1$ to $100~\Msun$.

After a star particle forms, it emits ionizing photons that ionize the
surrounding gas particles. We compute a spherical region that balances the
ionization and recombination using the Str\"omgren sphere model
\citep{Hopkins+2012, Renaud+2013, Baba+2017, Hirai+2017Rprocess, Fujii+2021b}.
All gas particles within the Str\"omgren sphere are heated to $10000~\mathrm{K}$
to mimic the emergence of H$_{\rm II}$ regions. The total number of ionizing
photons as a function of time is computed using {\tt P\'EGASE}
\citep{FiocRocca-Volmerange1997}.

Energy feedback and chemical evolution are handled by the latest version of the
chemical evolution library {\tt CELib} \citep{Saitoh2017} {\footnote
{\url{https://bitbucket.org/tsaitoh/celib/src/develop}}}(see
section~\ref{sec:models}). We have implemented a stellar-wind model in {\tt
CELib}, which computes the IMF-weighted mass loss of each star (simple stellar
population; hereafter SSP) particle at every time step and releases it to
surrounding $128\pm8$ gas particles without injecting energy or momentum.

For core-collapse supernovae (CCSNe) and Type Ia SNe, we implement feedback
models based on the momentum feedback approach used in FIRE-2 simulations
\citep{Hopkins+2018}. Feedback from CCSNe occurs at a random time between the
$N_\mathrm{CCSN}$th and $(N_\mathrm{CCSN}+1)$th SNe of each SSP particle by
integrating the IMF. The amount of released elements is computed during that
mass range. This procedure continues until the mass range reaches the lower
limit of the CCSNe mass range.

Type Ia SNe feedback follows a delay time distribution with a power law index of
$-1.1$ and a normalization of $1.3\times 10^{-3}$ events per unit solar mass
over $10~\mathrm{Gyr}$ \citep{MaozGraur2017}. We divide the cumulative DTD into
intervals with an expected count equal to unity and trigger one Type Ia SN event
at a uniformly random time within each interval. The ejecta and momentum are
injected into the surrounding gas particles similar to the CCSNe.

For AGB stars, we adopt the feedback model of \citet{Saitoh2017}. Based on age
tables, we integrate the AGB mass loss over IMF for stars in an SSP particle
that reach the end of their lifetimes between $i_{\rm AGB} \times
10^8~\mathrm{yr}$ and $(i_{\rm AGB} +1) \times 10^8~\mathrm{yr}$, and release
this at a randomly selected time within that interval.  Hence, the relatively
massive stars contribute the pollution first and the lower mass stars contribute
the feedback later. Neither energy nor momentum is injected into the surrounding
gas particles in the AGB models.

Metal mixing is solved using a shear-based turbulence model \citep{Shen+2010,
Saitoh2017} with a mixing coefficient of 0.01, which reproduces realistic
distributions of both $\alpha$ and $r$-process elements observed in the local
Universe \citep{HiraiSaitoh2017}. Further details about our chemical evolution
models are provided in section~\ref{sec:models}.

\subsection{Models} \label{sec:models}

Our simulations include four chemical evolution channels: stellar winds, CCSNe,
Type Ia SNe, and AGBs. All these models are implemented in {\tt CELib} and use
them from {\tt ASURA}. Here we solve the chemical evolution of eighteen elements
(H, He, C, N, O, Ne, Na, Mg, Si, P, S, Ca, Fe, Ni, Zn, Sr, Ba, and Eu). The five
elements, Na, P, Zn, Sr, and Ba, are added in the original implementation of
{\tt CELib}.

For stellar winds and CCSNe, we adopt the yield table of rotating massive stars
provided by \citet{LimongiChieffi2018} \footnote{\url{http://orfeo.iaps.inaf.it/}}
. The yield table includes time-dependent yield induced by stellar winds as well
as CCSNe and they cover the stellar masses from $10~\Msun$ to $120~\Msun$, the
initial metallicity [Fe/H] = $-3, -2, -1, 0$, and initial rotation velocities of
0, 150, and $300~{\mathrm{km}~{\mathrm{s}}^{-1}}$.  The metallicity-dependent
fraction of the stellar rotation velocity is taken from \citet{Prantzos+2018}. We
note that only stars with masses between $13$--$25~\Msun$ explode as CCSNe, while
more massive stars collapse directly to black holes (BHs).

As described above, we switch the IMFs depending on the local metallicity. Since
the yield table of \citet{LimongiChieffi2018} does not cover the very high-mass
and very low-metallicity ranges, we switch to another table. For very
low-metallicity CCSNe, i.e., $Z \le 10^{-5} \Zsun$, we employ the yield table of
\citet{Nomoto+2013} with the Susa IMF \citep{Susa2013}. All materials are
released at the timing of CCSNe, meaning that there are no stellar winds at this
metallicity range.

The stellar yields of Type Ia SNe are the metallicity-dependent ones of
\citet{Seitenzahl+2013}. In particular, we used the model N100. The delay-time
distribution follows a power law of $-1.1$ (e.g., \cite{Totani+2008,
MaozMannucci2012}). The event rate for 10 Gyr per $1~\Msun$ is set to 0.0013
which is consistent with the observational estimates (e.g., \cite{Maoz+2012}.
The offset of the event rate is $500~\mathrm{Myr}$ based on chemical evolution
models and observations of Local Group dwarf galaxies \citep{Homma+2015}.

The mass loss from AGB stars is also taken into account.  For our
fiducial model, we adopt the FRANEC Repository of Updated Isotopic Tables \&
Yields (FRUITY) to obtain AGB nucleosynthetic yields and associated mass return
\citep{Cristallo+2009, Cristallo+2011, Cristallo+2015}, whose tables cover the
mass range of $1.3$--$6.0~\Msun$. We also test an alternative set of yield
tables generated by the Monash University stellar evolution program
\citep{Karakas2010, Doherty+2014, CampbellLattanzio2008, Gil-Pons+2013}, which
covers the mass range of $1.0$--$9.0~\Msun$, to assess the sensitivity of
our results to the choice of AGB yield tables. We refer to this set of yield
tables as the Monash model.

Since the two stellar evolution codes are independent, they predict different
yields. Notable differences include the strength of the third dredge-up and the
hot-bottom burning processes. We note that these processes change the surface
compositions of stars significantly (e.g., \cite{KarakasLattanzio2014}). The
Monash models reach approximately three times higher temperatures at the bottom
of its convective envelope than that in FRUITY, resulting in different CNO cycle
products \citep{Cristallo+2015}. In particular, the Monash yields predict more
nitrogen and less carbon compared to the FRUITY.

As we noted above, we adopt the FRUITY yield tables as the fiducial model, and
the differences between the two sets of yield tables will be discussed in
section~\ref{sec:AGB}.

\begin{figure}[htb]
 \begin{center}
  \includegraphics[width=8cm]{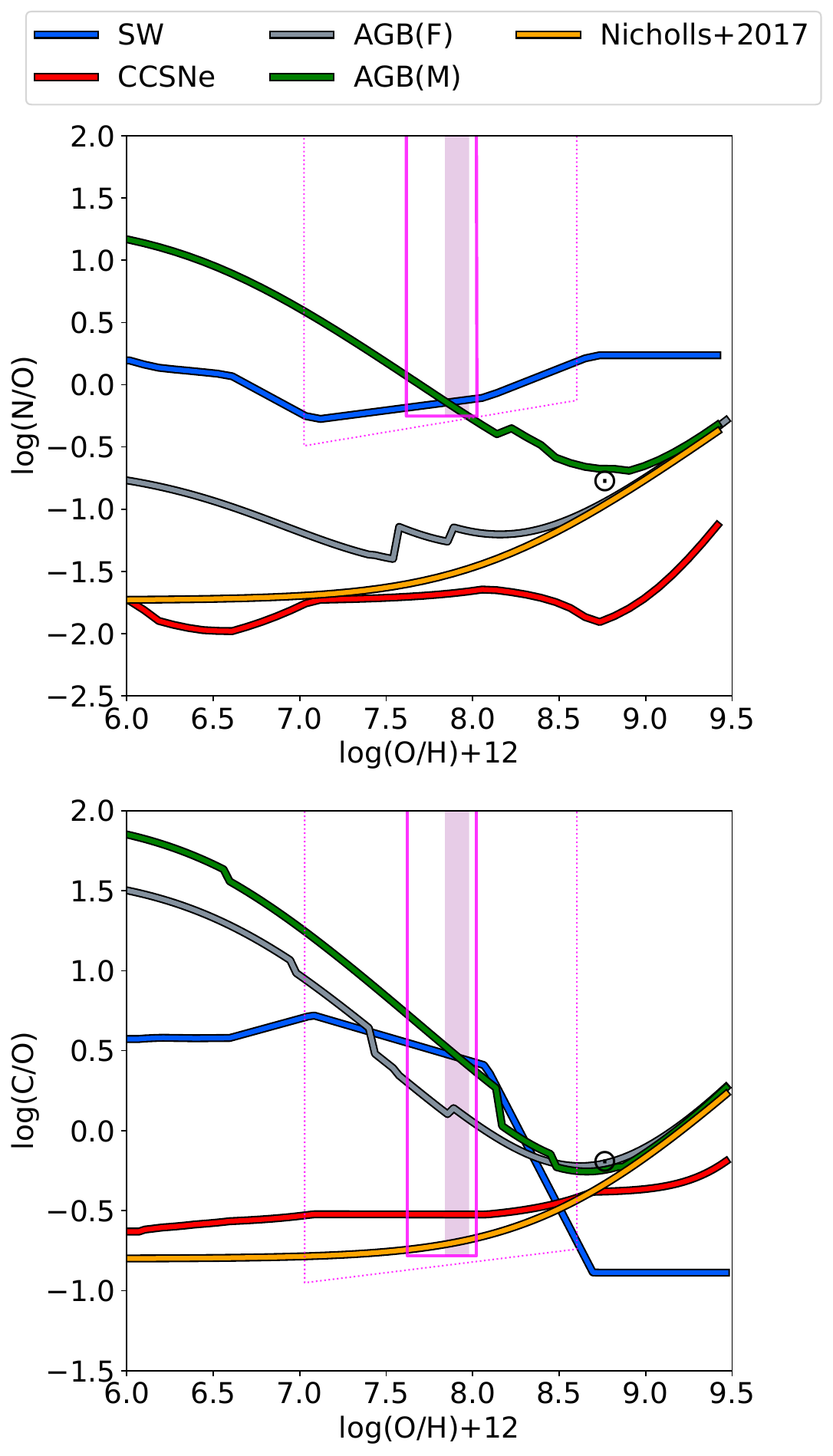}
 \end{center}
    \caption{Expected log(N/O) (top panel) and log(C/O) (bottom panel) as a
    function of metallicity, $\log({\rm O/H})+12$. The blue, red, gray, and green
    curves exhibit the IMF-weighted log(N/O) and log(C/O) of the ejecta for
    stellar wind (SW), CCSNe, AGBs (FRUITY yields), and AGBs (Monash yields),
    respectively. For AGBs, the ejecta components are diluted by the same mass
    of the gas component of the SSP particle. The orange curve represents the
    scaling values of log(N/O) and log(C/O) obtained from \citet{Nicholls+2017}.
    The fiducial and conservative regions of GN-z11 from \citet{Cameron+2023} are
    shown with magenta solid and dotted lines. The magenta filled region is
    the observational values of GN-z11 \citep{Alvarez-Marquez+2025}.
    The solar abundance values are shown with the solar symbols.
    {Alt text: Two-panel line plot showing nitrogen-to-oxygen (top) and
    carbon-to-oxygen (bottom) ratios against oxygen abundance. Multiple curves
    represent theoretical yields from stellar winds, core-collapse supernovae,
    and two asymptotic giant branch star models. The curves demonstrate that
    stellar winds and asymptotic giant branch stars are possible origins of high
    nitrogen-to-oxygen ratios. Observational scaling relations and the measured
    values of GN-z11 are also shown for comparison.}}
    \label{fig:YieldsComparison}
\end{figure}

Figure~\ref{fig:YieldsComparison} shows the comparison of the yields for log(N/O)
and log(C/O) as a function of the metallicity for the primary three channels in
our adopted models. Note that the Type Ia SNe are negligible owing to their long
delay times in this study.

We adopt the solar abundance pattern from \citet{Asplund+2009}. Thus, the solar
values are $\log(\mathrm{N/O}) _{\odot} = -0.86$, $\log(\mathrm{C/O}) _{\odot} =
-0.26$, and $\log(\mathrm{O/H})_{\odot}+12 = 8.69$, respectively.

Table \ref{tab:model} summarizes the models used in this study. We refer to the
model that uses the Limongi \& Chieffi yields for the stellar wind and CCSNe,
includes the Pop III mode at $Z<10^{-5}\Zsun$, employs the FRUITY AGB yields,
and adopts the normal LWB intensity of \citet{Incatasciato+2023} as our fiducial
model.

To study the contributions of the LWB, we ran a model which turns off LWB
($f_\mathrm{LWB}$ = 0; NoLWB). To examine the contributions of the Pop III mode,
we conducted a run where the Pop III mode is turned off (NoPopIII).  We also
studied the impacts of the top-heavy IMF (TopHeavy) and the contributions of the
AGB mass loss (NoAGB). Additionally, we switched the AGB yield table from FRUITY
to Monash (Monash). These comparisons were performed by turning off a single
function and comparing the results with the fiducial model.

We mainly focus on the fiducial model in section~\ref{sec:results}. We then
discuss the differences among models in section~\ref{sec:comparisons}.

\begin{table*}
\tbl{Simulation models used in this study\label{tab:model}}{%
\begin{tabular}{lcccccl}
\hline
Model & LWB & Pop III mode & Pop II/I IMF & AGB mass loss & AGB yield & Description \\
\hline
Fiducial & \ding{51} & \ding{51} & Chabrier & \ding{51} & FRUITY  & Fiducial model with all processes \\
NoLWB    & $\times$   & \ding{51} & Chabrier & \ding{51} & FRUITY  & Turn off LWB \\
NoPopIII & \ding{51} & $\times$   & Chabrier & \ding{51} & FRUITY  & Turn off PopIII \\
TopHeavy & \ding{51} & \ding{51} & Top heavy& \ding{51} & FRUITY  & Adopt a top-heavy IMF ($\propto m^{-0.35}$)\\
NoAGB    & \ding{51} & \ding{51} & Chabrier & $\times$   & $\times$& Turn off AGB mass loss \\
Monash   & \ding{51} & \ding{51} & Chabrier & \ding{51} & Monash  & Monash AGB yields instead of FRUITY \\
\hline
\end{tabular}}
\end{table*}

\section{Results} \label{sec:results}

\subsection{Evolution of star-forming halos until $z\sim10$} \label{sec:starforming_halo}

\begin{figure}[htb]
 \begin{center}
  \includegraphics[width=8cm]{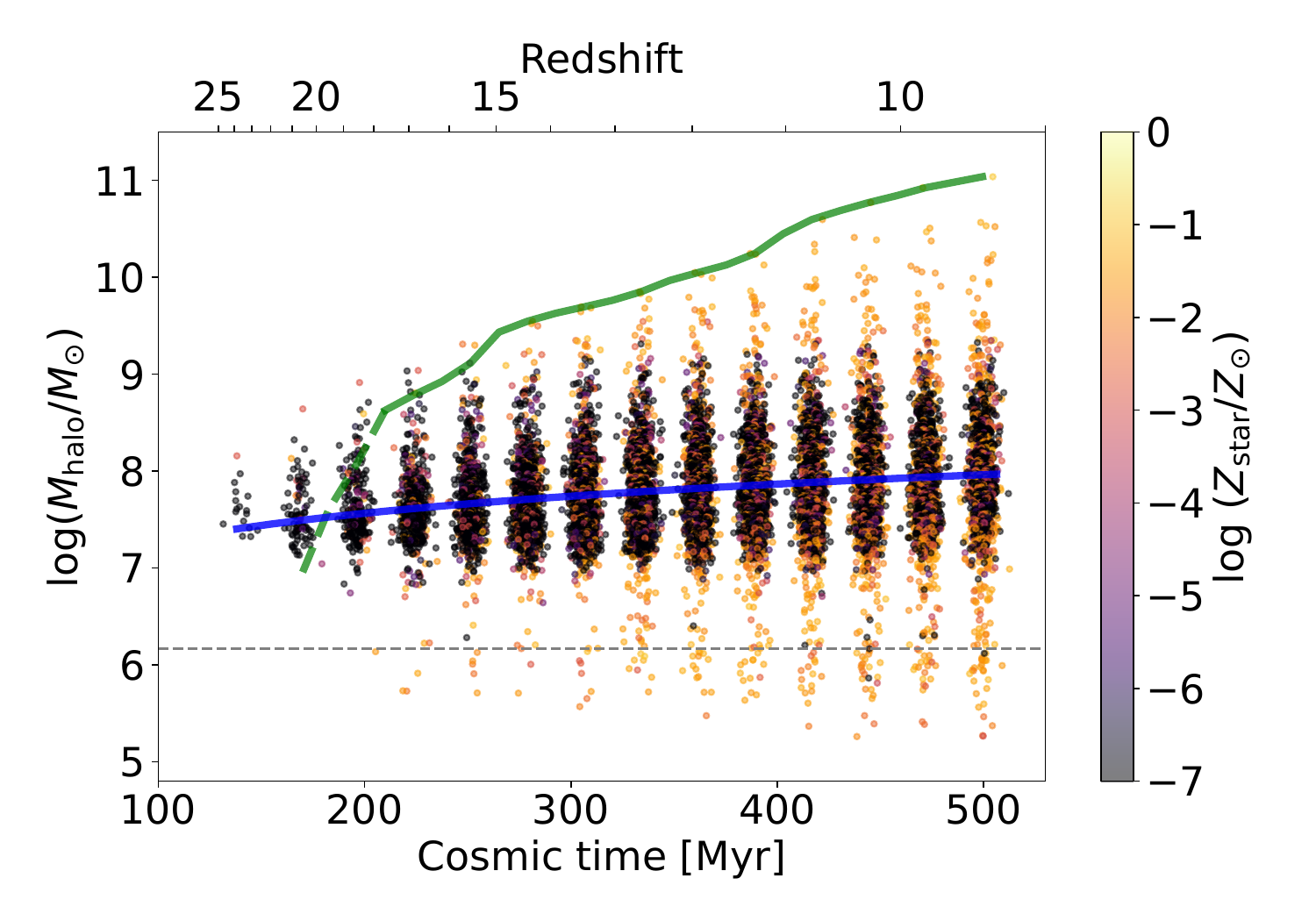}
 \end{center}
    \caption{Distribution of star-forming halos as a function of time (redshift).
    Each dot represents the mass of the halo at the redshift.  Color
    indicates the average metallicity of the stellar component in each halo.
    For visibility, we add a spread for time. Each group is sampled at the
    same redshift. Here we sample every $\sim 27.6~\mathrm{Myr}$.  The blue curve
    exhibits the expected halo mass of the virial temperature at $10^4~\mathrm{K}$
    obtained from \citet{BrommYoshida2011}. The green curve is the evolution
    track of the primary halo we investigate in this study. The solid part has
    stars while the dashed part does not.
    {Alt text: Scatter plot showing halo mass versus time and redshift. Dots
    represent individual halos with metallicity indicated for each halo. A reference
    curve shows expected halo mass for virial temperature of ten to the fourth
    Kelvin. An evolution track represents the primary halo evolution, with solid
    and dashed segments indicating presence or absence of stars.}
    }
\label{fig:HaloMassStellarZ}
\end{figure}

First, we show the evolution of the halos in the computational domain.  Here we
take into account all the halos found in the zoom-in region. Halos contaminated
by the boundary particles are not considered. Figure~\ref{fig:HaloMassStellarZ}
shows the distribution of star-forming halos from $t = 141~\mathrm{Myr};z=23.6$
to $500~\mathrm{Myr};z=9.60$. The definition of star-forming halos is that
they have stars.  We can see that the mass of the star-forming halos is
typically $> 10^{7} ~\Msun$ in the early phase and spreads the distribution from
$10^{4} ~\Msun$ to $10^{11}~\Msun$ around $z=10$. The median of the distribution
follows the expected halo mass of the virial temperature of $10^4~\mathrm{K}$
obtained from \citet{BrommYoshida2011}.

A mass of $\sim 10^{7}~\Msun$ or more is required to host stars for the
very low-metallicity halos (represented by black circles) due to the inefficient
cooling of the primordial gas. Although there are some variations, our results
match the theoretical expectation (the blue curve). From this figure, we can
also see that higher metallicity halos are more likely to have larger variations
in masses. For the massive halos, this reflects the self-enrichment owing to
chemical evolution in galaxies, while for the less massive halos, metal polluted
gas can allow star formation even in shallower potentials.

Our target halo is the most massive halo found in this figure at the final
epoch.  We track the main branch of this halo using {\tt AHF}. The progenitor
was detected at $t\sim180~\mathrm{Myr};z\sim20$, and star formation in it began
at $t\sim220~\mathrm{Myr};z\sim17$. By $t\sim270~\mathrm{Myr};z\sim15$, it had
become the most massive halo in the zoom-in region. From here, we focus on the
evolution, structure, and chemical composition of the halo.

The total, gas, and stellar masses of the primary halo are $3.3\times10^{10}
~\Msun$, $3.1\times10^{9}~\Msun$, and $1.2\times10^{9}
~\Msun$, respectively, at $t=431~\mathrm{Myr};z=10.70$. Its virial
radius is $\sim 10~{\rm kpc}$ at this epoch. These become $5.7\times10^{10}
~\Msun$, $6.07\times10^{9}~\Msun$, $9.6\times10^{9}
~\Msun$, and $\sim12~\mathrm{kpc}$ at $t=472~\mathrm{Myr};z=10.00$.  We
note that the halo mass of $3.3\times10^{10}~\Msun$ at
$t=431~\mathrm{Myr}; z=10.70$ is consistent with that of GN-z11 obtained by
\citet{Scholtz+2024} from an extrapolation of stellar mass-halo mass relation
\citep{Behroozi+2013}. We also note that \citet{Ferrara2024} expects that the
halo mass is $10^{11.5}~\Msun$ based on the halo abundance matching of
the luminosity function.

\subsection{Evolution of the primary galaxy until $z\sim10$}\label{sec:evolution}

\begin{figure*}[tp]
 \begin{center}
  \includegraphics[width=14cm]{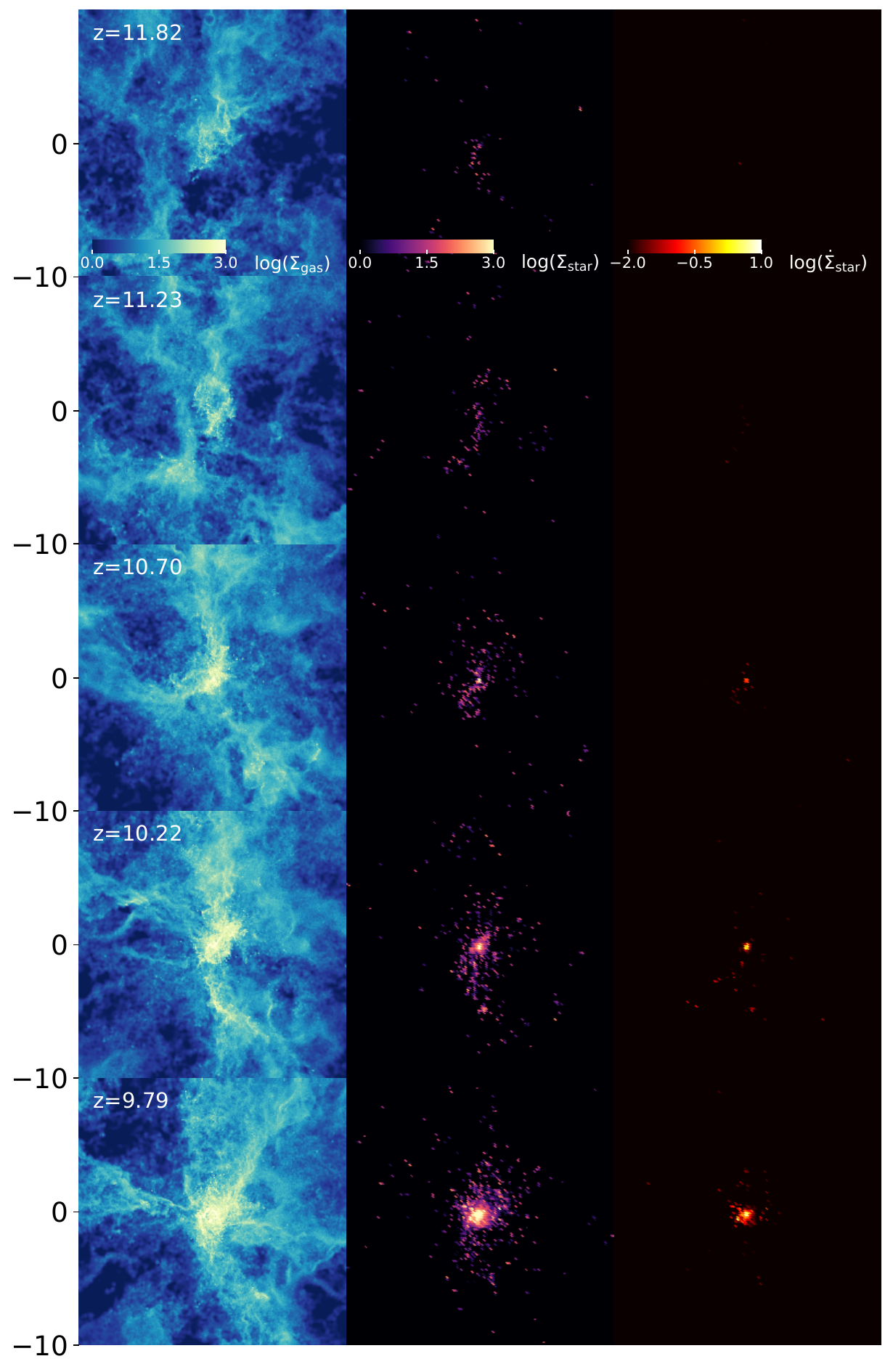}
 \end{center}
    \caption{Left column: The projected gas surface density in the units of
    $\Msun~\mathrm{pc}^{-2}$. Colors represent the surface density and
    the color bar is found in the top panels. Middle column: The projected
    stellar density in the units of $\Msun~\mathrm{pc}^{-2}$. Right
    column: the surface SFR density in the units of $\Msun~\mathrm{yr}
    ^{-1}~\mathrm{kpc}^{-2}$. Each panel covers the $20~\mathrm{kpc} \times
    20~\mathrm{kpc}$ region. From top panels to bottom panels, we show the
    snapshots from $z=11.82$ to $z=9.8$.
    {Alt text: Three-column panel showing projected densities across a twenty
    kiloparsec by twenty kiloparsec region. Panels arranged from top to bottom
    show snapshots at redshifts from eleven point eight two to nine point seven
    nine. Left column displays gas surface density in solar masses per square
    parsec. Middle column shows stellar density in solar masses per square
    parsec. Right column presents star formation rate surface density in solar
    masses per year per square kiloparsec. The central high-gas density region
    is the intersection of filamentary structures. Star-forming regions and
    stars are highly concentrated at the center.}
    }
\label{fig:20kpc}
\end{figure*}

\begin{figure*}[tp]
 \begin{center}
  \includegraphics[width=14cm]{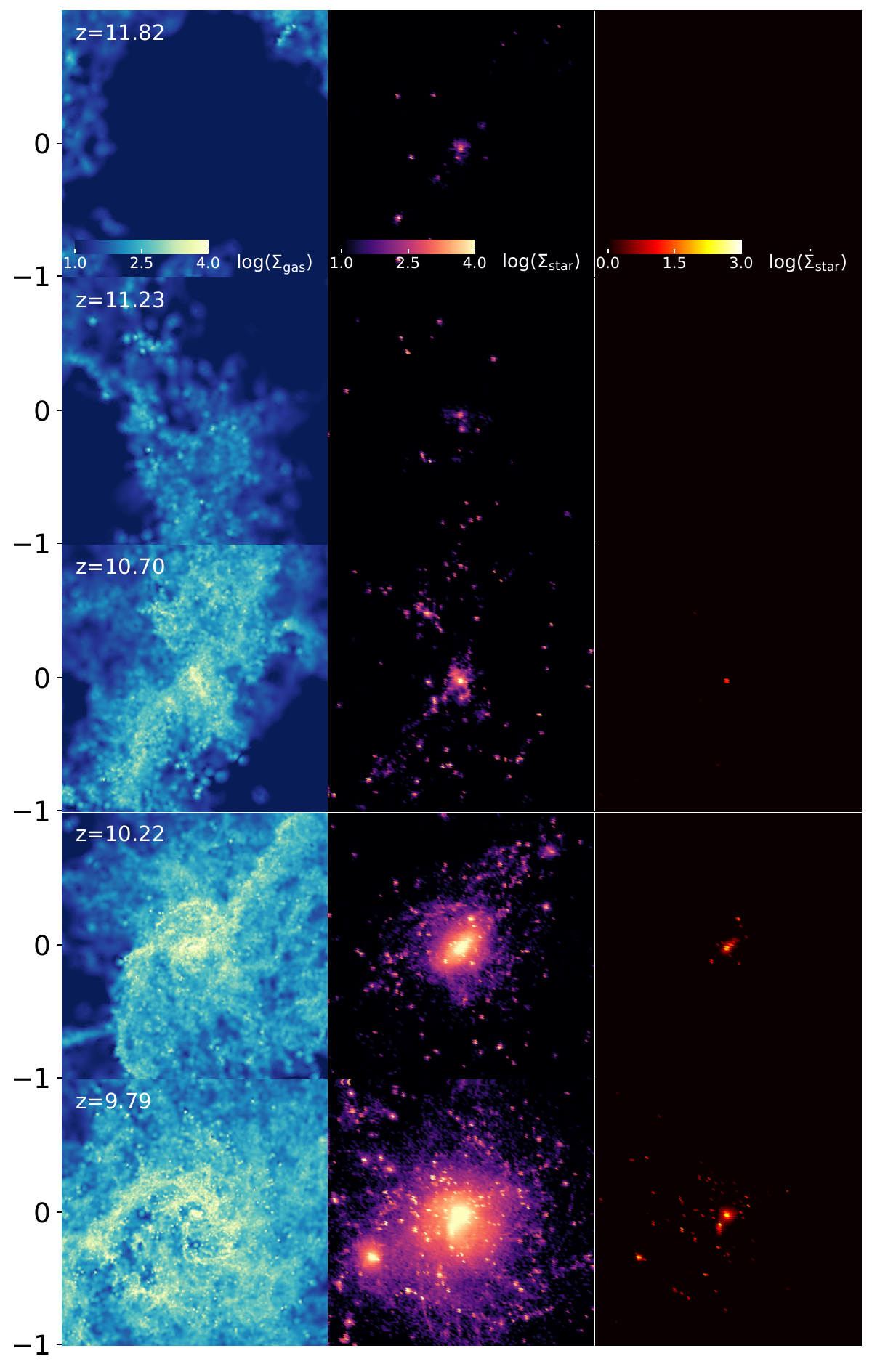}
 \end{center}
    \caption{Same as figure~\ref{fig:20kpc}, but for the $2~\mathrm{kpc} \times
    2~\mathrm{kpc}$ regions.
    {Alt text: Three-column panel showing projected densities across a two
    kiloparsec by two kiloparsec region. Panels arranged from top to bottom show
    snapshots at redshifts from eleven point eight two to nine point seven nine.
    Left column displays gas surface density in solar masses per square parsec.
    Middle column shows stellar density in solar masses per square parsec. Right
    column presents star formation rate surface density in solar masses per year
    per square kiloparsec. Gas accumulates at the center, forming a very compact
    and dense galaxy.}
    }
\label{fig:2kpc}
\end{figure*}

Figures~\ref{fig:20kpc} and \ref{fig:2kpc} show the global evolution of surface
gas density ($\Msun~\mathrm{pc}^{-2}$), surface stellar density
($\Msun~\mathrm{pc}^{-2}$), and surface SFR density ($\Msun
~\mathrm{yr}^{-1}~\mathrm{kpc}^{-2}$) around the most massive halo. The
figures depict these properties at scales of $20~\mathrm{kpc} $ and
$2~\mathrm{kpc}$ on a side, respectively. Figure~\ref{fig:20kpc} includes
almost the entire halo.

In figure~\ref{fig:20kpc}, a distinct knot at the center, located at the
intersection of large-scale filaments, is visible. At $t>420~\mathrm{Myr};z<11$,
three or four prominent gas filaments connect from the knot in the center to the
outer region, leading to large-scale gas accretion. Stars are distributed
sparsely in the early phase. By $t=431~\mathrm{Myr};z=10.70$, a dense stellar
clump forms in the center and continues to evolve in later phases. The halo also
contains high-density peaks in the stellar mass distributions, identified as
star clusters. The surface SFR density reveals rapid star formation occurring
predominantly in the center.

Further detailed structural evolutions are shown in figure~\ref{fig:2kpc}.
At the first epoch ($t=375~\mathrm{Myr};z=11.82$), there is a void in the gas
distribution with a radius of $\sim 1~\mathrm{kpc}$ in the center.  A tiny
galaxy is visible in the corresponding stellar distribution. The void structure
in the gas distribution is likely the relic of the early burst. After this brief
break (a short drop, not a full suspension of star formation), continuous gas
accretion onto the center resumes along the filaments, forming a compact
rotating gas disk with arms at the center. The surface SFR density at the center
is $10^3~\Msun~\mathrm{yr}^{-1}~\mathrm{kpc}^{-2}$, which is as high as
those of local star-forming galaxies \citep{Genzel+2010}.

Star formation is highly concentrated even at this scale, with less prominent
star formation along the filaments. This suggests that a part of stellar clumps
originate from minihalos, brought in via accretion (the ex-situ origin). The
final stellar distribution exhibits rich structures, including the central
galaxy, stellar clumps with various sizes, tidal features, and a diffuse stellar
halo.

\begin{figure*}[tp]
 \begin{center}
  \includegraphics[width=15cm]{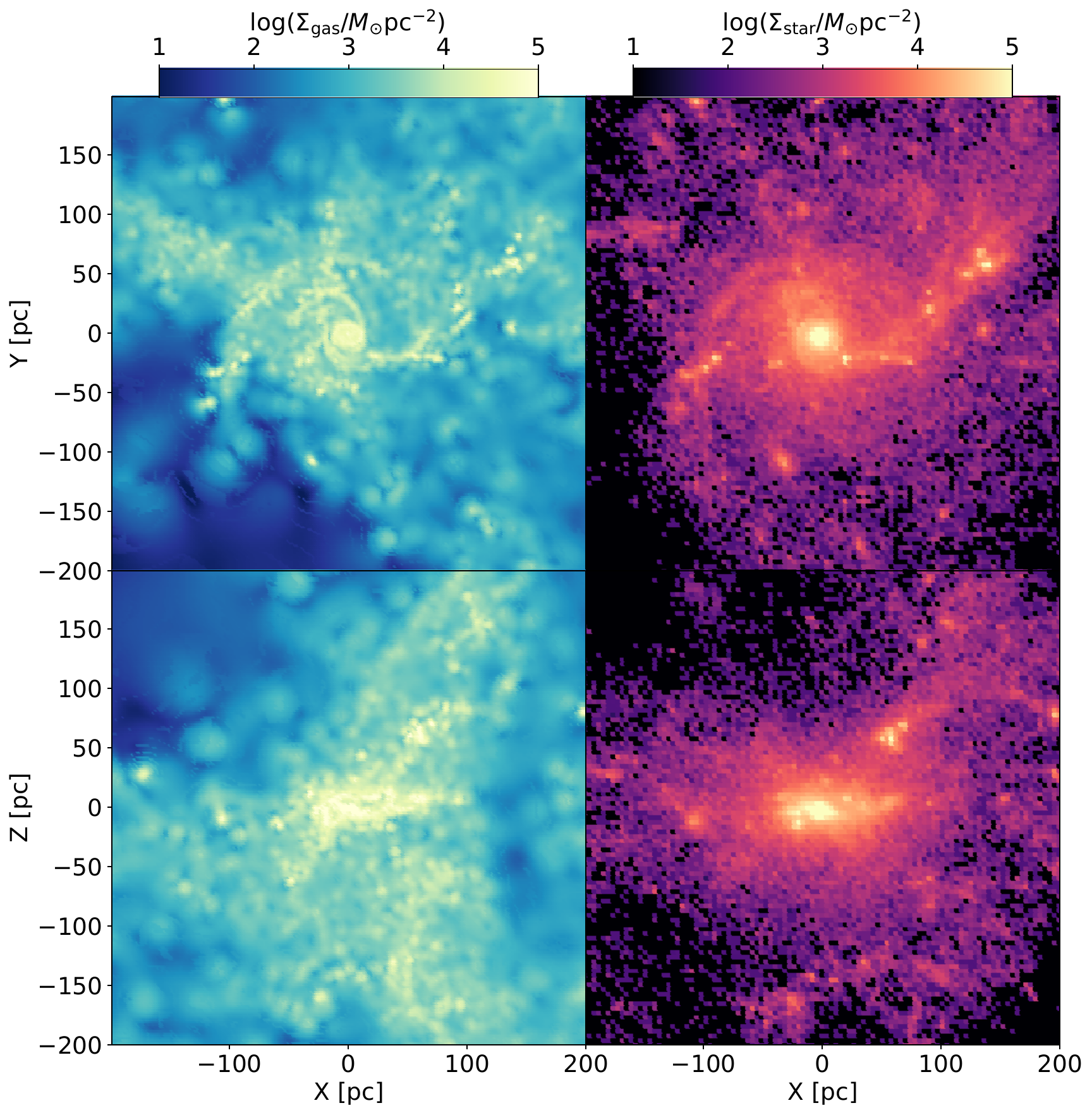}
 \end{center}
    \caption{Gas and stellar surface density maps at $z=10.46$. Left panels:
    Projected surface density maps of gas in units of $\Msun~\mathrm{pc}^{-2}$.
    Right panels: Projected surface density maps of stars in units of
    $\Msun~\mathrm{pc}^{-2}$. The orientations have been adjusted to
    present face-on and edge-on views in the top and bottom panels,
    respectively.
    {Alt text: Two-column panel showing surface density maps across a four
    hundred parsec by four hundred parsec region. Left column shows gas
    surface density in solar masses per square parsec. Right column displays
    stellar surface density in solar masses per square parsec. The maps show
    formation of a compact, rotation-supported galaxy.}
    }
\label{fig:200pc}
\end{figure*}

Figure~\ref{fig:200pc} shows the close-up views of the galaxy at
$t=444~\mathrm{Myr};z=10.46$, where the star formation is in its most active
epoch. The face-on view of the gas surface distribution reveals a compact
condensation of gas with arm-like structures. The edge-on view displays a thin
gas distribution with a radius of $100~\mathrm{pc}$. A chain of gas clumps in
the first quadrant in the edge-on views, which also seen in the face-on view,
suggests gas accretion from various directions, as seen in the global views of
evolution in figures~\ref{fig:20kpc} and \ref{fig:2kpc}.

The surface stellar density maps (the right panels of figure~\ref{fig:200pc})
show a strong concentration of stars at the center. The majority of stars are
confined within a radius of several tens of parsecs. The structure forms a
slightly extended disk. While dense clumps of stars surround the central
galaxy, they are less prominent and few in number---approximately a dozen are
visible in the viewing region.

This structure resembles a disk galaxy, but with a very compact size of several
$10~\mathrm{pc}$. Recent observations using JWST/ALMA have confirmed the
existence of such high-$z$ disk galaxies (e.g., \cite{Li+2023, Fujimoto+2025,
Rowland+2024, Scholtz+2025}). Current simulations targeting this era are also
able to reproduce high-$z$ compact galaxies (see also \cite{Pallottini+2022,
Yajima+2023}).

\begin{figure*}[tp]
 \begin{center}
  \includegraphics[width=15cm]{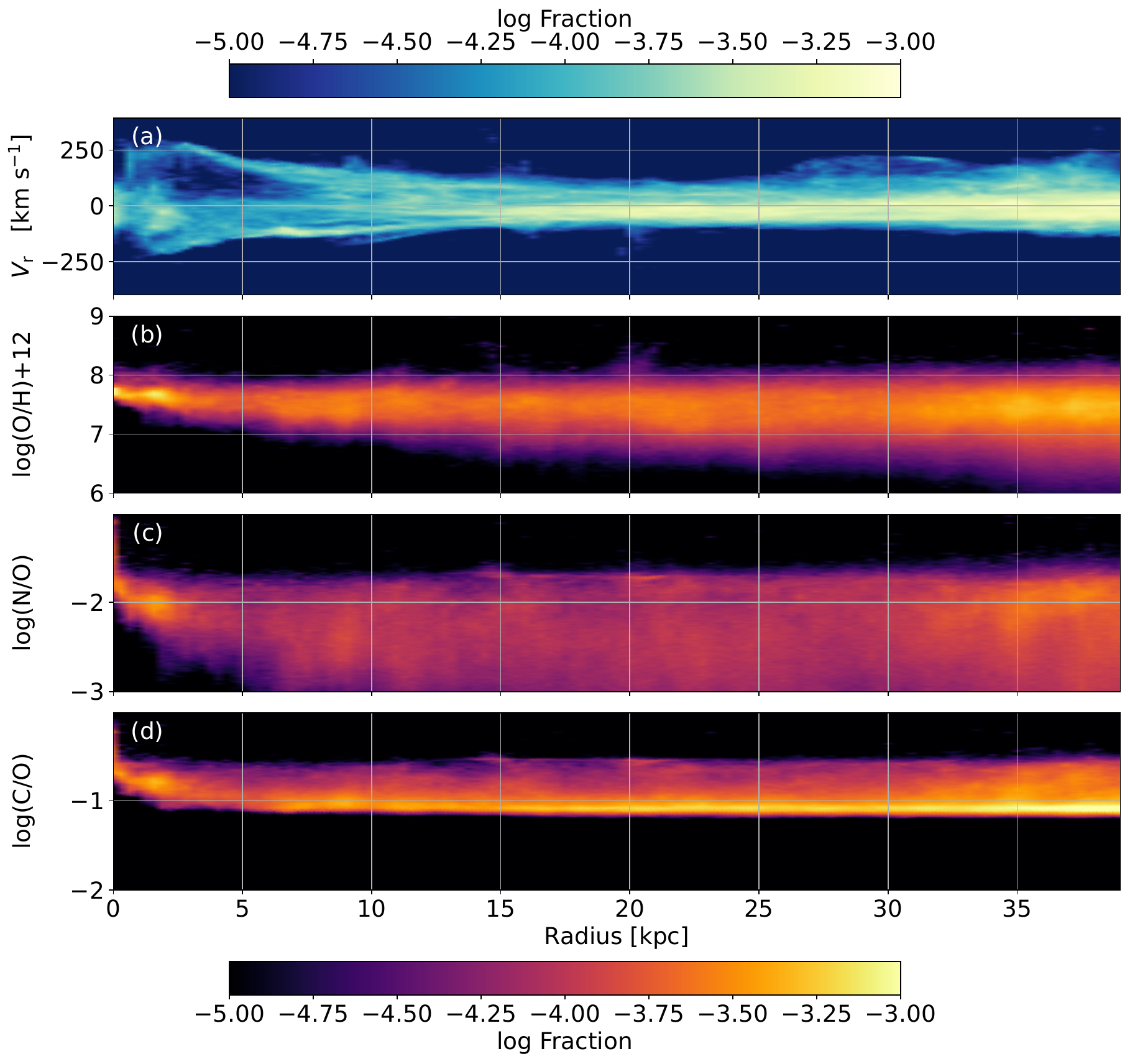}
 \end{center}
    \caption{(a) Radial velocity, (b) metallicity, (c) log(N/O), and (d) log(C/O)
    profiles of gas from the galactic center to $40~\mathrm{kpc}$. The snapshot of
    $t=444~\mathrm{Myr};z= 10.46$ is used.
    {Alt text: Four-panel plot showing radial distributions from galactic
    center to forty kiloparsecs. Panels display radial velocity,
    oxygen-abundance, nitrogen-to-oxygen ratio, and carbon-to-oxygen ratio as
    functions of distance. Radial velocity decreases gradually from outer to
    inner, and finally reaches minus two hundred fifty kilometers per second.
    Metallicity and abundance ratios are relatively flat in the outer regions,
    while they gradually increase at the inner region.}
    }
\label{fig:phase:Rvel_OH12}
\end{figure*}

The radial velocity profile of gas up to $40~\mathrm{kpc}$ ($\sim 4\times
R_\mathrm{vir}$) is shown in panel (a) of figure~\ref{fig:phase:Rvel_OH12}.
The negative $V_\mathrm{r}$ component from the center to more than
$\sim20~{\rm kpc}$ indicates that there is a large-scale accretion with
$V_\mathrm{r} \sim 100~{\mathrm{km}~{\mathrm{s}}^{-1}}$. The dissipation due to
shocks is crucial for making a compact system.  While it is less prominent, we
can see the positive $V_\mathrm{r}$ component within $R_\mathrm{vir}$, which
likely traces outflows or leaked high-speed accretion with an offset from the
galactic center.

\subsubsection{Star formation history}

\begin{figure}[htb]
 \begin{center}
  \includegraphics[width=8cm]{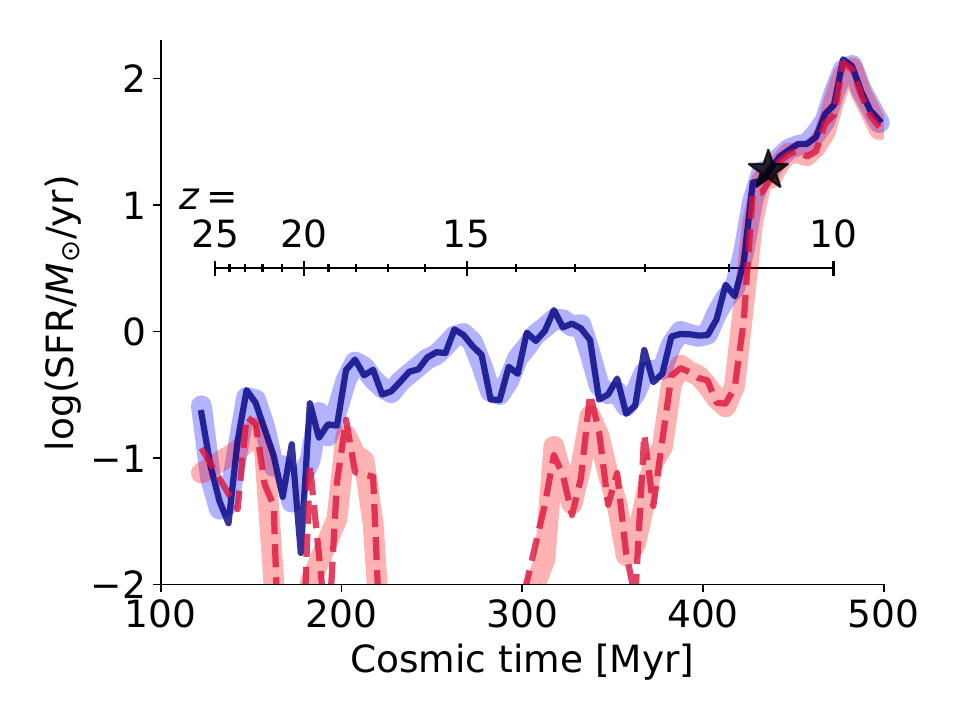}
 \end{center}
    \caption{SFRs as a function of the cosmic age and redshift. The blue solid
    curve represents the star formation history of all the stars in the primary
    halo at $t \sim 500~\mathrm{Myr};z\sim9.64$. The red, dashed curve shows the
    SFR of the central $1~\mathrm{kpc}$ of star particles in the primary halo.
    Thick blue and red curves are the smoothed SFRs by the Savitzky-Golay method
    \citep{SavitzkyGolay1964}. The star symbol indicates the value of GN-z11
    \citep{Bunker+2023}.
    {Alt text: Line plot showing star formation rate versus cosmic time and
    redshift. Two curves display the entire galaxy and the central one
    kiloparsec region. At redshifts below eleven, star formation rate in the
    central region dominates the overall star formation rate of the galaxy. The
    central star formation rate is almost comparable with that of GN-z11.}
    }
\label{fig:SFR}
\end{figure}

Figure~\ref{fig:SFR} shows the star formation history of the target halo.  In
the early phase, the SFR is below unity, with several weak bursts. Around
$t=420~\mathrm{Myr};z=11$, it rapidly increases, and reaches a maximum of $>
100~{\Msun~\mathrm{yr}^{-1}}$. Then, it decreases. This ``excited
phase'' of star formation, starburst, is characterized by intense activity in
the central region of the primary halo (see figure~\ref{fig:2kpc}). The duration
of the starburst phase is less than $100~\mathrm{Myr}$. The bursty star
formation and rapid quenching of it are frequently predicted in JWST confirmed
high-$z$ galaxies (e.g., \cite{Tacchella+2023,Looser+2025, Robertson+2024JOF,
Topping+2024}).

The comparison between the blue and red curves clearly demonstrates that the
starburst phase is highly concentrated in the central region. This centrally
confined starburst is triggered by the large-scale gas accretion through
filaments, as shown in figure~\ref{fig:20kpc}. This accreted gas brings angular
momentum, which leads to the formation of the disk-like structure.

\begin{figure}[htb]
 \begin{center}
  \includegraphics[width=8cm]{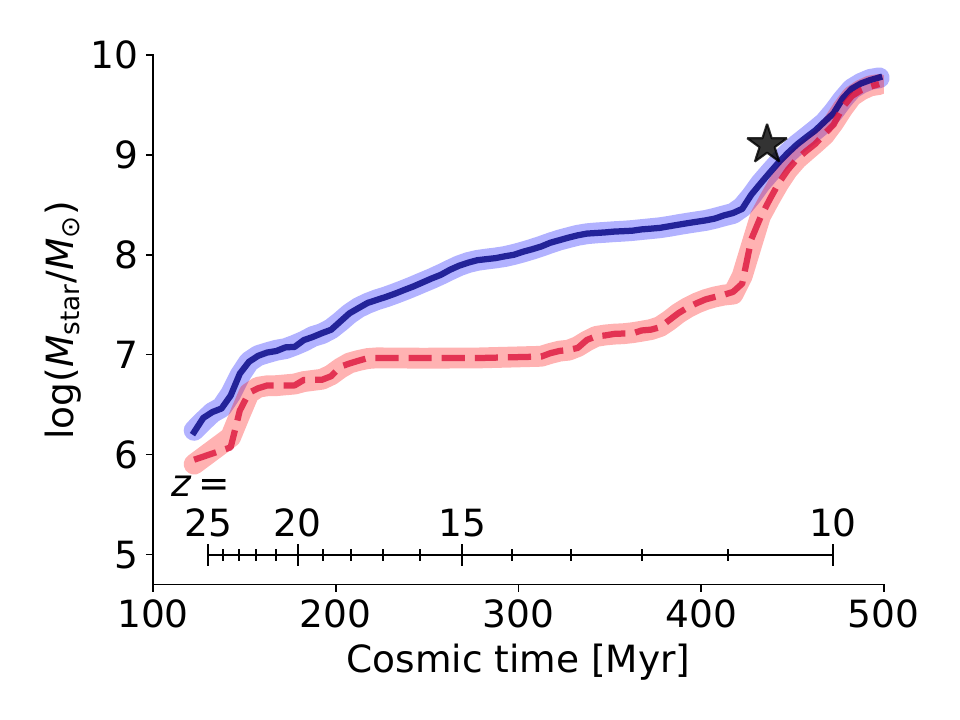}
 \end{center}
    \caption{Mass evolutions of the primary halo and within their central
    $1~\mathrm{kpc}$. The blue solid curve represents the mass evolution of the
    all stars in the primary halo at $t=500~\mathrm{Myr};z=9.64$ whereas the red dashed curve
    indicates that within $1~\mathrm{kpc}$. The star symbol indicates the value
    of GN-z11 \citep{Bunker+2023}.
    {Alt text: Line plot showing stellar mass versus cosmic time and redshift.
    Two curves display the entire galaxy and the central one kiloparsec region.
    Below redshift eleven, the central stellar mass dominates. At redshift
    around ten point five, the stellar mass is comparable to that of GN-z11.}
    }
\label{fig:MassEvolution}
\end{figure}

The mass evolution of the primary halo, which corresponds to the star formation
history, is shown in figure~\ref{fig:MassEvolution}.  The stellar mass undergoes
rapid growth at $t>420~\mathrm{Myr};z<11$, primarily at the central region
($R<1~\mathrm{kpc}$). During this phase, the mass of the central region increases
by over two orders of magnitude. At $t=440~\mathrm{Myr};z\sim 10.5$, the stellar
mass becomes comparable to that of GN-z11 \citep{Tacchella+2023}. Notably, at
this point, the SFR reaches several tens of $\Msun~\mathrm{yr}^{-1}$,
which is also comparable to that of GN-z11. Consequently, the galaxy at this
epoch serves as an excellent analog to GN-z11.

\begin{figure}[htb]
 \begin{center}
  \includegraphics[width=8cm]{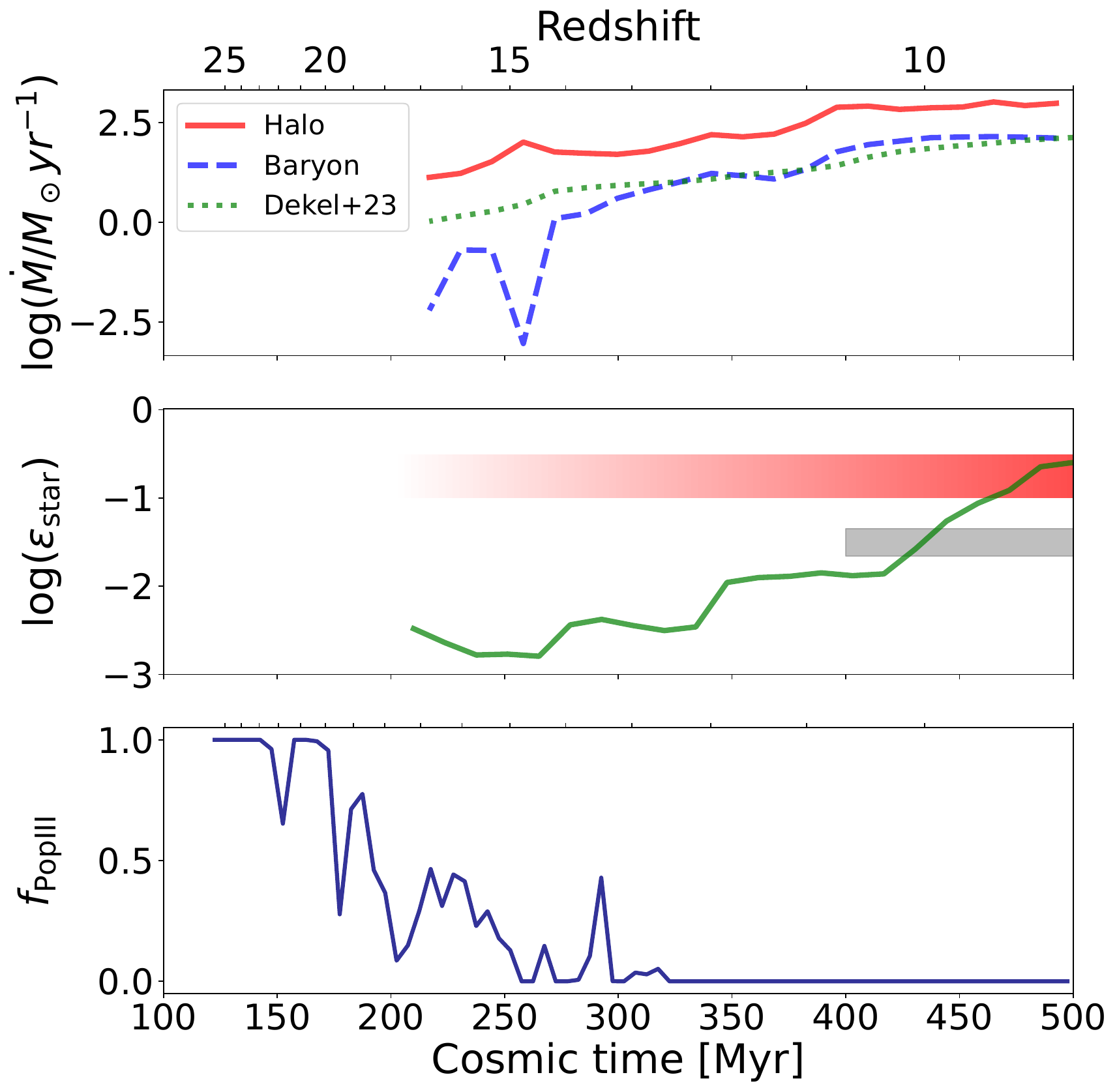}
 \end{center}
    \caption{Top panel: the halo (the red solid curve) and baryon (the blue
    dashed curve) accretion rates as a function of time and redshift.
    The green dotted curve is the mean baryonic accretion rate on to the halo
    evaluated using Eq. 31 in \citet{Dekel+2023}.
    Middle panel: the integrated galaxy formation efficiency as a function
    of the time and redshift (green curve). The grey shaded region represents
    the galaxy formation efficiencies of the halo mass of $3.3 \times 10^{10}
    ~\Msun$ and $5.7 \times 10^{10}~\Msun$. The efficiencies
    are computed by using equation 7 in \citet{Stefanon+2021}.  The red shaded
    region represents the star formation efficiencies of $0.14$--$0.3$ suggested
    by the abundance-matching technique \citep{Inayoshi+2022}.  Bottom panel:
    The Pop III fraction of the forming stars at a given time and redshift. The
    mass-based fraction is adopted.
    {Alt text: Line plot showing time evolution of three key quantities in three
    panels. Top panel displays halo and baryon mass accretion rates versus time.
    The baryon accretion rate is initially below the cosmic mean but converges
    with it after redshift thirteen. Middle panel shows star formation
    efficiency starting small, then rapidly increasing to match
    abundance-matching predictions near the end of the simulation. Bottom panel
    presents the population three star fraction over time, which is nearly unity
    at earliest times and decreases steeply to negligible levels after redshift
    thirteen.}
    }
\label{fig:efficiencies}
\end{figure}

The top panel of figure~\ref{fig:efficiencies} shows the halo and baryon
accretion rates of the last $250~\mathrm{Myr}$ until the cosmic time of
$500~\mathrm{Myr}$ ($z\sim9$). As in figure~\ref{fig:HaloMassStellarZ}, we track
only the main branch of the merger tree to reconstruct this figure. The mass
accretion rate of the halo (i.e., the total mass) starts at $\sim
10~\Msun~\mathrm{yr^{-1}}$ and reaches $\sim 10^3~\Msun~\mathrm{yr^{-1}}$ in the
end. This rapid growth supports the formation of a $10^{10}~\Msun$ halo.
Initially, the baryon accretion rate is lower because of mass leakage from the
progenitor and accreting halos due to feedback.  It quickly recovers and follows
the halo accretion rate with the cosmic baryon fraction of $0.16
(\sim\Omega_\mathrm{b}/\Omega_\mathrm{M} \equiv f_\mathrm{b}) $. Between
$t=420~\mathrm{Myr};z=11$ and $t=470~\mathrm{Myr};z=10$, the accretion rate is
$\sim 10^2~\Msun~\mathrm{yr^{-1}}$, suggesting that a substantial fraction of
the global accretion material is funneling into the galactic center.  This
concentration results from the relatively insufficient development of tidal
fields at rare halos, the selective infall of low-angular momentum gas
(segregation), and cancellation of angular momentum at the galactic center
\citep{Dubois+2012}.

This accretion is key to understanding the establishment of early starburst
galaxies. Due to the low angular momentum, the centrally concentrated starburst
with inefficient feedback is triggered (see figure~\ref{fig:SFR}). Continuous
accretion induces a situation of the feedback-free starburst \citep{Dekel+2023}.

Following \citet{Ceverino+2024}, we depict the integrated galaxy formation
efficiency, $\epsilon_{\rm star}$ of the primary halo in the middle panel of
figure~\ref{fig:efficiencies}. The definition is
\begin{equation}
    \epsilon_{\mathrm{star}} = M_\mathrm{star}/(f_\mathrm{b} M_\mathrm{halo}).
\end{equation}
The efficiency starts with a low value of $\log_{10}(\epsilon_\mathrm{star})
\sim -2.5$ from the time when the first stars are detected in the main
progenitor ($t=210~\mathrm{Myr}$; $z\sim18$). The efficiency gradually increases
and rapid growth is observed at $t<420~\mathrm{Myr};z>11$, where the starburst
takes place (see figure~\ref{fig:SFR}).

The analytical model of the feedback-free starburst \citep{Li+2024FFB,
Dekel+2023}, the abundance matching technique \citep{Inayoshi+2022}, and
numerical simulations of galaxy formation (e.g., \cite{Yajima+2023,
Ceverino+2024}) predict high efficiencies in high-$z$ galaxies. Our simulation
is consistent with these studies. The maximum efficiency obtained in the final
epoch of our simulation is $\sim 0.25 (\log_{10}(\epsilon_\mathrm{star}) \sim
-0.6)$, almost an order of magnitude higher than the observational data expected
before JWST era \citep{Stefanon+2021}, while very close to the analysis based on
JWST observations \citep{Inayoshi+2022}.  It is also in agreement with the
feedback-free starburst model \citep{Li+2024FFB}.

The Pop III fraction is also shown in the bottom panel of
figure~\ref{fig:efficiencies}. Almost all the stars formed in the main
progenitor are not Pop III stars, as shown in \citet{Yajima+2023}. In our model,
strong feedback from Pop III is available only in minihalos and becomes
negligible at $t > 300~\mathrm{Myr}$;$z < 14$. This shift in star formation
makes it difficult to halt gas accretion, yielding a high galaxy formation
efficiency.

\subsubsection{Stellar mass-SFR and stellar mass-metallicity relations}

\begin{figure}[htb]
 \begin{center}
  \includegraphics[width=8cm]{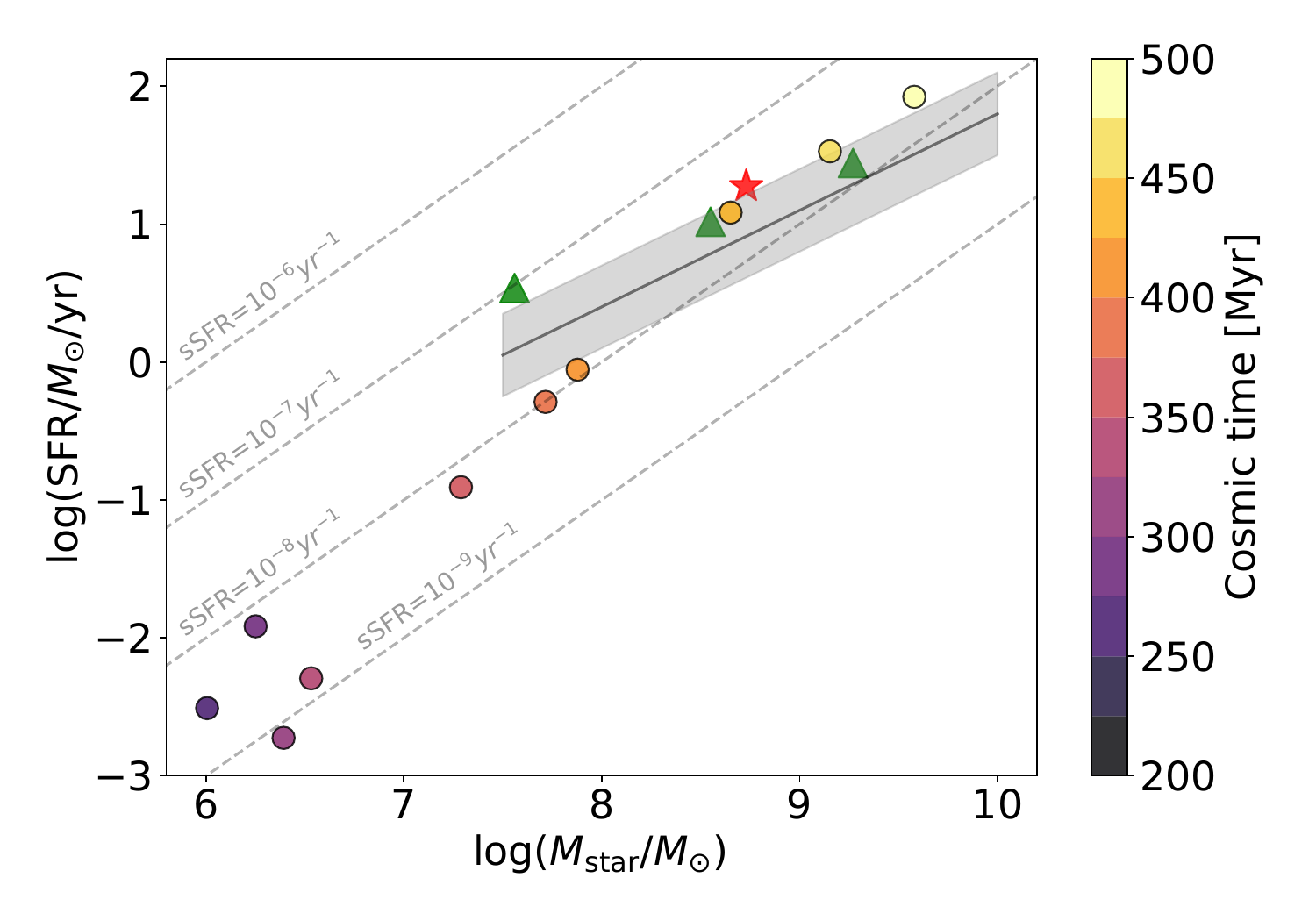}
 \end{center}
    \caption{The relationship between stellar mass and SFR in the primary halo
    from $t=251~\mathrm{Myr};z=15.75$ to $t=500~\mathrm{Myr};z=9.60$.  The
    circles represent the data points of each epoch and the color indicates the
    corresponding time in Myr. The interval of the data points is
    $27.6~\mathrm{Myr}$.  The red star represents the value of GN-z11
    \citep{Bunker+2023}.  The green triangles show the averaged values of three
    representative mass ranges for samples between $z=4$ and $10$
    \citep{Nakajima+2023}. The black line with the gray shaded region shows the
    distribution of the star-forming galaxies at $z=7$--$10$ \citep{Curti+2024}.
    The gray dotted lines give the constant specific SFRs.
    {Alt text: Scatter plot displaying star formation rate versus stellar mass.
    Circles represent the target galaxy at different epochs with time indicated
    for each data point. Values around redshifts eleven to ten closely match
    observed measurements of GN-z11 and other high-redshift galaxies. Dotted
    lines indicate constant specific star formation rates.}
    }
\label{fig:MstarSFR}
\end{figure}

Figure~\ref{fig:MstarSFR} shows the evolution of the stellar mass and SFR of the
galaxy. In the early phase ($t\lesssim360~\mathrm{Myr};z\gtrsim12$), the
specific SFR (sSFR) remains low, between $10^{-9}$ and $10^{-8}~\mathrm{yr}^{-1}$.
Then, the sSFR increases by $\sim0.5$ dex at $t=444~\mathrm{Myr};z=10.46$,
becoming comparable to that of GN-z11 (the red star; \cite{Bunker+2023}).  At
this point, the sSFR is also in good agreement with the average sSFRs for
samples between $z=4$ and $10$ (blue stars; \cite{Nakajima+2023}) and
star-forming galaxies at $z=7$--$10$ (the black line with grey shaded
region; \cite{Curti+2024}). This high sSFR corresponds to a rapid increase in SFR at
the center of the galaxy (figures~\ref{fig:SFR} and \ref{fig:MassEvolution}).
In the last two epochs, this galaxy exceeds the observations.  However, the star
formation is being quenched, and we suspect that it will return to the fiducial
range later through dilution by accreted gas.

\begin{figure}[htb]
 \begin{center}
  \includegraphics[width=8cm]{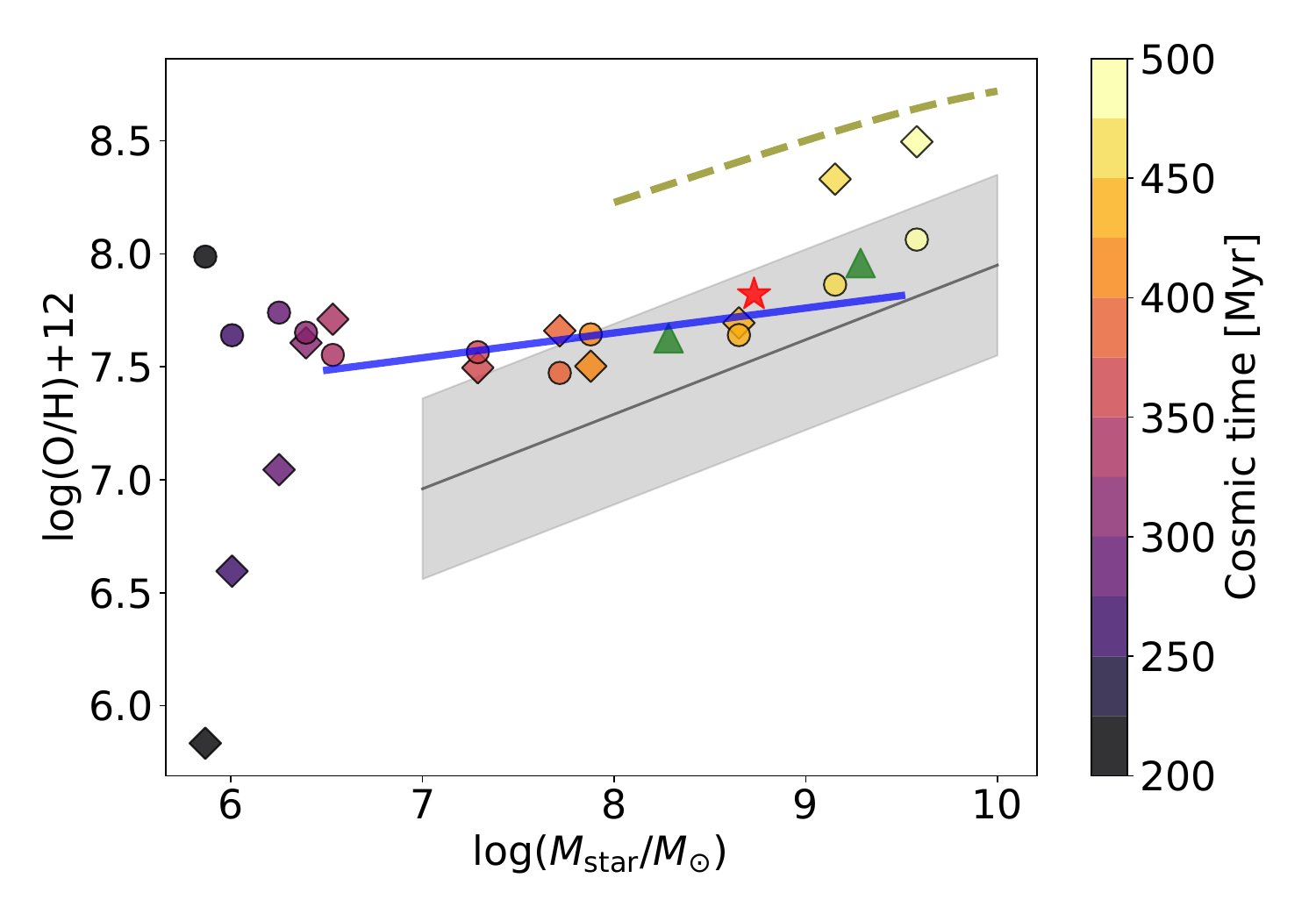}
 \end{center}
    \caption{The stellar mass-metallicity relation in the primary halo from
    $t=251~\mathrm{Myr};z=15.75$ to $t=500~\mathrm{Myr};z=9.60$.  Circles and
    diamonds represent the gas and stallar components respectively, with colors
    indicating the corresponding time in Myr. The red star represents the value
    of GN-z11 \citep{Bunker+2023}. The green triangles show the averaged values
    of two mass ranges for samples between $z=8$ and $10$ \citep{Nakajima+2023}.
    The blue line is the mass-metallicity relation for $z=6$--$10$
    \citep{Curti+2024}.  The black line with the gray shaded region shows the
    distribution of the star-forming galaxies at $z=7$--$10$ \citep{Heintz+2023}.
    The olive dashed curve represents the median of mass-metallicity relation
    at $z=0$ \citep{Curti+2020}.
    {Alt text: Scatter plot displaying metallicity versus stellar mass. Circles
    and diamonds represent gas-phase and stellar-phase metallicities of the
    target galaxy at different epochs respectively. Values closely match
    observed measurements of GN-z11 (star symbol) and other high-redshift
    galaxies (triangles and shaded region). Stellar metallicities at final
    epochs approach those of local galaxies.}
    }
\label{fig:MstarZ}
\end{figure}

Figure~\ref{fig:MstarZ} illustrates the evolution of stellar mass and
metallicity [$\log$(O/H)+12]. Initially the gas-phase metallicity (circles)
starts at a high value of $>8$ due to pollution by stellar feedback with a
non-negligible amount of Pop III stars. The metallicity then weakly dilutes and
keeps around $7.6$ for $\sim 100~\mathrm{Myr}$. This is owing to the dilution
by continuous accretion of low-metallicity intergalactic gas. At
$t=444~\mathrm{Myr};z=10.46$, the metallicity value becomes close to that of
GN-z11 (the red star; \cite{Bunker+2023}). Metallicities at these epochs align
well with observations \citep{Nakajima+2023,Heintz+2023, Curti+2024}.
Interestingly, the flat part closely matches observations of high-$z$ galaxies
around $z=6$--$10$ \citep{Curti+2024} and far away from those of the local
Universe \citep{Curti+2020}. Shortly after, it rises slightly, even though the
value is consistent with the observations \citep{Heintz+2023}.

The figure also plots the stellar metallicity, $\log$(O/H)+12, with diamond
symbols. The stellar metallicity starts from a very low value, rapidly increases
to $\sim 7.6$, and remains unchanged for the next $\sim 100~\mathrm{Myr}$.
During this plateau phase, the metallicities of the stellar and gas components
are almost identical, indicating a well mixed situation. As the starburst begins,
the metallicity increases rapidly, resulting in a significant offset from
high-$z$ observations and approaching the observations of the local Universe
\citep{Curti+2020}. This strong self-enrichment results from highly concentrated
star formation in the central region, coupled with relatively weak feedback
(i.e., no Pop III stars).  Notably, while star-forming galaxies at $z\sim3$
typically show lower stellar metallicities than gas-phase ones
\citep{Sommariva+2012}, our simulation at $t> 472~\mathrm{Myr};z<10$ shows the
opposite pattern. This suggests that the simulated distribution is a limited
case and would be relaxed in lower-$z$ under more mild star formation.

Observations of extremely metal-poor galaxies show that the gas-phase
metallicity has a lower limit of $Z=0.01 \Zsun$, i.e., the metallicity floor
(e.g., \cite{Wise+2012, Nishigaki+2023}).  Throughout the entire evolution
from $t=251~\mathrm{Myr};z=15.75$, the metallicity of the galaxy remains above
$0.1 \Zsun$. The stellar metallicity starts below this floor but quickly becomes
enriched.

In panel (b) of figure~\ref{fig:phase:Rvel_OH12}, we plot the metallicity
distribution from the galactic center to the intergalactic medium up to $\sim 4
\times R_\mathrm{vir}$. The metallicity of gas surrounding the galaxy reaches
$0.1 \Zsun$ with a certain variation. Due to the continuous accretion along the
filaments, the averaged galactic metallicity remains at $\sim 0.1 \Zsun$ on a
longer timescale.

\subsubsection{Radial profiles}

\begin{figure*}[htb]
 \begin{center}
  \includegraphics[width=16cm]{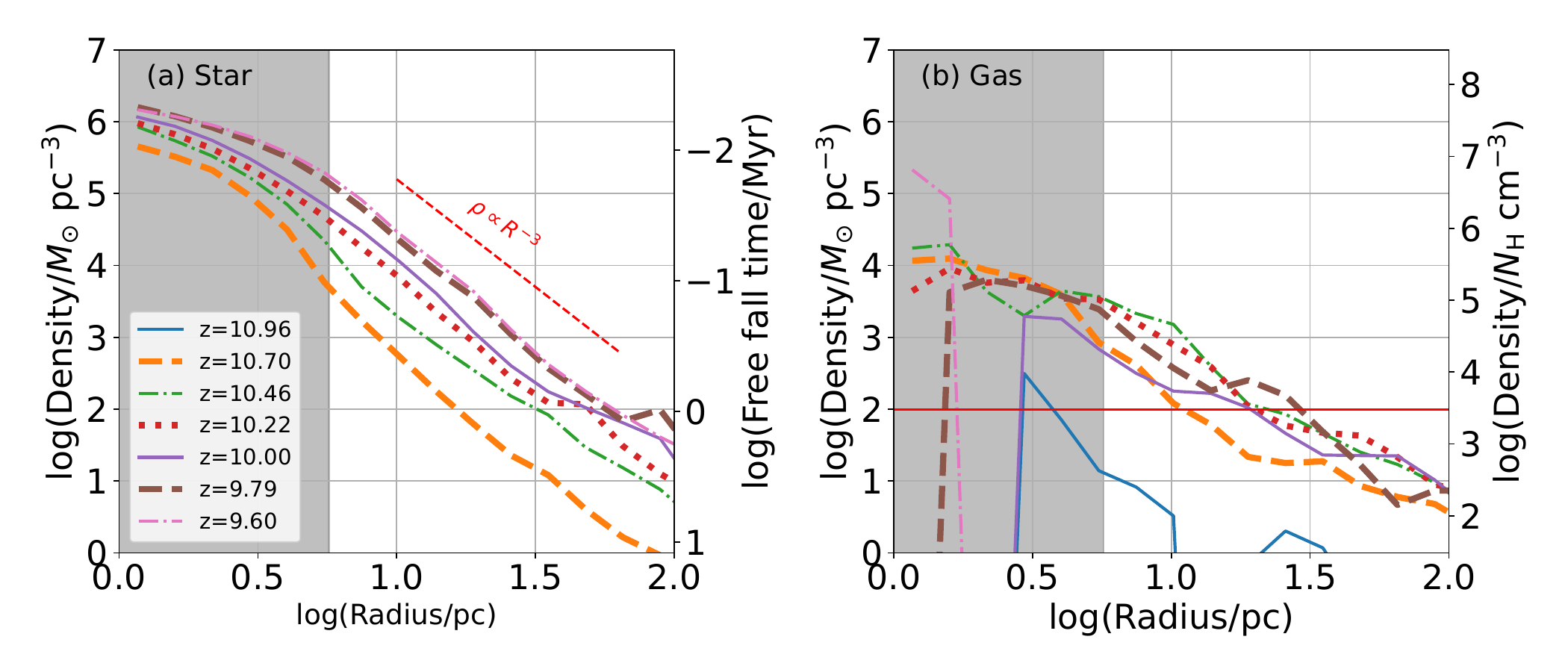}
 \end{center}
    \caption{Evolution of the radial density profiles every $13.8~\mathrm{Myr}$ from
    $t=431~\mathrm{Myr};z=10.70$ to $500~\mathrm{Myr};z=9.60$. Panel (a):
    Stellar radial profiles. The earliest profile is not shown because of out of
    range. The red dashed line represents the density which is inversely
    proportional to the cube of the radius. The gray region is less than the
    adopted softening length at $z\le9$. The vertical axis of the right-hand
    side represents the corresponding free fall time in units of Myr. Panel (b):
    Gaseous radial profiles. The vertical axis of the right-hand side shows the
    hydrogen number density. The red horizontal line exhibits the critical
    density for the feedback-free starburst predicted by \citet{Dekel+2023}.
    {Alt text: Line plot showing radial density profiles of stars and gas at
    multiple epochs. Stellar densities peak at the center and follow the
    inverse-cube power law at large radii. Gaseous density profiles are highly
    time-dependent. The horizontal line indicates the critical density of the
    feedback-free starburst and inner region exceeds this threshold density.}
    }
\label{fig:RadialProfile}
\end{figure*}

Panel (a) of figure~\ref{fig:RadialProfile} illustrates the evolution of the
stellar radial profile. This figure reveals a rapid evolution in the central
region of the galaxy. During the initial $27~\mathrm{Myr}$
($t=431$--$458~\mathrm{Myr}$;$z=10.70$--$10.22$), the profile increases by
approximately one order of magnitude, reflecting bursty star formation.
Subsequently, the evolution rate nearly stagnates in the final two epochs.

The central density is comparable to the young star clusters observed in the
local Universe (e.g., \cite{StolteBrandner2005,SelmanMelnick2013,
SharmaRodriguez2025}), dense GCs \citep{BaumgardtHilker2018},  and simulated
young star clusters \citep{Fujii+2024}. This extreme density could potentially
lead to the formation of supermassive stars via collisions, resulting in
intermediate-mass BH formation \citep{Fujii+2024}.

It should be noted that the density profiles are softened due to the limited
spatial resolution, in turn, the actual profile would be denser. The density
profile in the outer part is proportional to $R^{-3}$. Such a profile aligns
with observations of local GCs (e.g., \cite{Carballo-Bello+2012}).

While the radial profile of the stellar component evolves monotonically, that of
the gas component is more complicated, as expected. Panel (b) in
figure~\ref{fig:RadialProfile} shows the radial density profiles of the gas
component. At $t=417~\mathrm{Myr};z=10.96$, the profile does not extend to the
center, and the density is relatively low. Then, gas accretion toward the
galactic center occurs. The typical density within $10~\mathrm{pc}$ exceeds the
threshold density of the feedback-free starburst predicted by \citet{Dekel+2023}.
Consequently, the central gas rapidly converts into stars within a timescale
shorter than the typical CCSNe timescale ($\sim 10~\mathrm{Myr}$). This high
density continues until $t=472~\mathrm{Myr};z=10.00$. The burst observed in
figure~\ref{fig:SFR} is driven by this extremely high density. Afterward,
feedback regulates star formation by expelling gas from the galactic center.

Although our spatial resolution is insufficient, the central density reaches $>
10^4~\Msun~\mathrm{pc}^{-3}$, which overcomes the far-UV and extreme-UV
radiations, resulting in high-efficiency star formation
\citep{FukushimaYajima2022}. In our model, we implemented the H$_\mathrm{II}$
region model based on the Str\"omgren sphere \citep{Baba+2017,Fujii+2021b} as an
early feedback process. However, it does not affect the central evolution
because of the high central density.

The half-mass radius of the stellar component, $r_\mathrm{h}$, is $\sim
100~\mathrm{pc} $ and the enclosed mass in the radius is $M_\mathrm{star}
(<r_\mathrm{h}) \sim 6 \times 10^8~\Msun$ at $ t=431~\mathrm{Myr};
z=10.70$. The radius is almost comparable with that of GN-z11, which has
$r_\mathrm{h} \sim 100~\mathrm{pc}$ \citep{Tacchella+2023}. The succeeding
intense star formation builds up the central dense stellar component, resulting
in a compact stellar component of $\sim 80~\mathrm{pc}$ at $t=472~\mathrm{Myr};
z=10.00$. The enclosed mass at this epoch is almost an order of magnitude larger
than that at $t=431~\mathrm{Myr};z=10.70$. We note that observations of
high-$z$ galaxies by JWST report size-mass relations and the typical
half-mass/luminous radii of several hundreds pc (e.g., \cite{Yang+2022,
Ono+2023, Allen+2025, Westcott+2024, Robertson+2024JOF}). \citet{Harikane+2025}
show that high-$z$ galaxies ($z>6$) are categorized into two groups: one is
compact ($\lesssim 100~\mathrm{pc}$) and the other is extended (see their
figure~16). Our simulated galaxy associates with the first group in the sense of
size.

\begin{figure}[htb]
 \begin{center}
  \includegraphics[width=8cm]{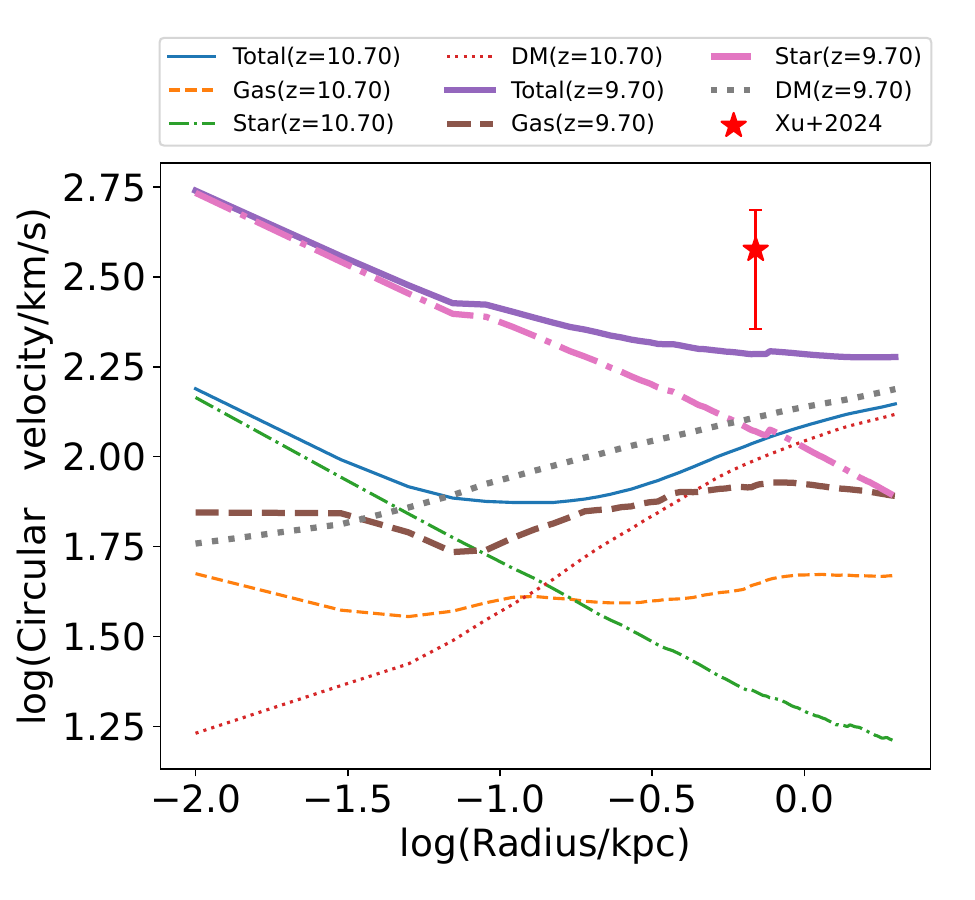}
 \end{center}
    \caption{Circular velocities as a function of radius for two different
    epochs, $z=10.70$ and $9.79$. The contributions of the gas, stellar, and
    dark matter components are shown. The observational rotation velocity
    obtained by \citet{Xu+2024} is expressed by a red star symbol with error
    bars.
    {Alt text: Line plot showing circular velocity versus radius at two epochs.
    Curves represent contributions from gas, stars, and dark matter components.
    Total circular velocity increases significantly between two epochs,
    particularly in the inner region. Final circular velocity is slightly lower
    than the observational measurement.}
    }
\label{fig:RC}
\end{figure}

We show the circular velocity of the galaxy in figure~\ref{fig:RC}. Here, we
compare two different epochs: before ($t=431~\mathrm{Myr};z=10.70$) and after
($t=486~\mathrm{Myr};z=9.79$) the burst. Before the burst, the circular velocity
curve has a downward convex profile.  The inner region of $R<20~\mathrm{pc}$ is
dominated by the stellar component, with a circular velocity of $V_\mathrm{c}
\sim 160~\mathrm{km}~{\mathrm{s}}^{-1} $ at $R=10~\mathrm{pc}$. Dark matter
dominates the outer part.

After the burst, the circular velocity increased dramatically, reflecting the
rapid growth of the central region. The value of $V_\mathrm{c}$ reaches $\sim
560~{\mathrm{km}~{\mathrm{s}}^{-1}}$ at $R=10~\mathrm{pc}$, indicating that the
central mass increased by more than an order of magnitude within
$R<10~\mathrm{pc}$. This rapid growth of the stellar component shifts the inner
edge where dark matter dominates from $R\sim160~\mathrm{pc}$ to
$\sim500~\mathrm{pc}$.

Combined with the visual inspection of the gas and stellar distributions
(figure~\ref{fig:200pc}) and the circular velocity profile
(figure~\ref{fig:RadialProfile}), we can conclude that this galaxy has a
rotation-dominated structure. In fact, our analysis shows that the fraction of
the rotation velocity in the two-dimensional plane is almost 60\% of the
circular velocity within $200~\mathrm{pc}$.
{\footnote {
We computed the fraction of the rotation velocity in the two-dimensional plane as
follows.  First, we selected all baryon particles within $200~\mathrm{pc}$ from the
galactic center.  For each particle, we calculated the two-dimensional radial
velocity:
\begin{equation}
    v_\mathrm{rad} = (x v_x+ y v_y)/\sqrt{x^2+y^2},
\end{equation}
where $x$, $y$, $v_x$, and $v_y$, are the position and velocity components of
the baryon particle. Here, the disk lies in the $x-y$ plane. The rotation
velocity can be measured as
\begin{equation}
    v_\mathrm{rot} = \sqrt{v_\mathrm{total}^2 - v_\mathrm{rad}^2},
\end{equation}
where $v_\mathrm{total} = \sqrt{v_x^2+v_y^2}$. From this, we obtain the rotation
fraction in the two-dimensional plane, $f_\mathrm{rot} = v_\mathrm{rot}
/v_\mathrm{total}$. Finally, we calculated the mass-weighted average of
$f_\mathrm{rot}$ for all selected particles.}}

\citet{Xu+2024} derive the rotation velocity of GN-z11. They report that the
rotation velocity at the half-light radius of the core is $376^{+110}_{-150}
~\mathrm{km}~{\mathrm{s}}^{-1}$, which is higher than the expected circular
velocity at the virial radius. This indicates a steeply increasing velocity
curve toward the inner region. Our simulated galaxy exhibits a similar trend in
its circular velocity curve, albeit with an even more centrally concentrated
structure. In our simulation, the rapidly rotating center forms due to intense
gas accretion toward the inner region.

We have shown the formation process of the primary galaxy up to
$t\sim470~\mathrm{Myr};z\sim10$ in this section. The central starburst begins at
$t\sim420~{\rm Myr};z\sim 11$ and lasts until $t\sim470~\mathrm{Myr};z\sim10$.
This starburst event leads to the formation of a massive, dense central stellar
component.  The SFR, the stellar mass, the $M_\mathrm{star}$-SFR relation and
the $M_\mathrm{star} $-metallicity relation are consistent with high-$z$
galaxies.  Notably, this galaxy resembles GN-z11.

\subsection{Nitrogen, Carbon, and Oxygen Abundances}\label{sec:chemicalcomposition}

In this section, we focus on the evolution of the chemical composition of the
primary galaxy, whose evolution and physical states resemble those of GN-z11.

\begin{figure}[htb]
 \begin{center}
  \includegraphics[width=8cm]{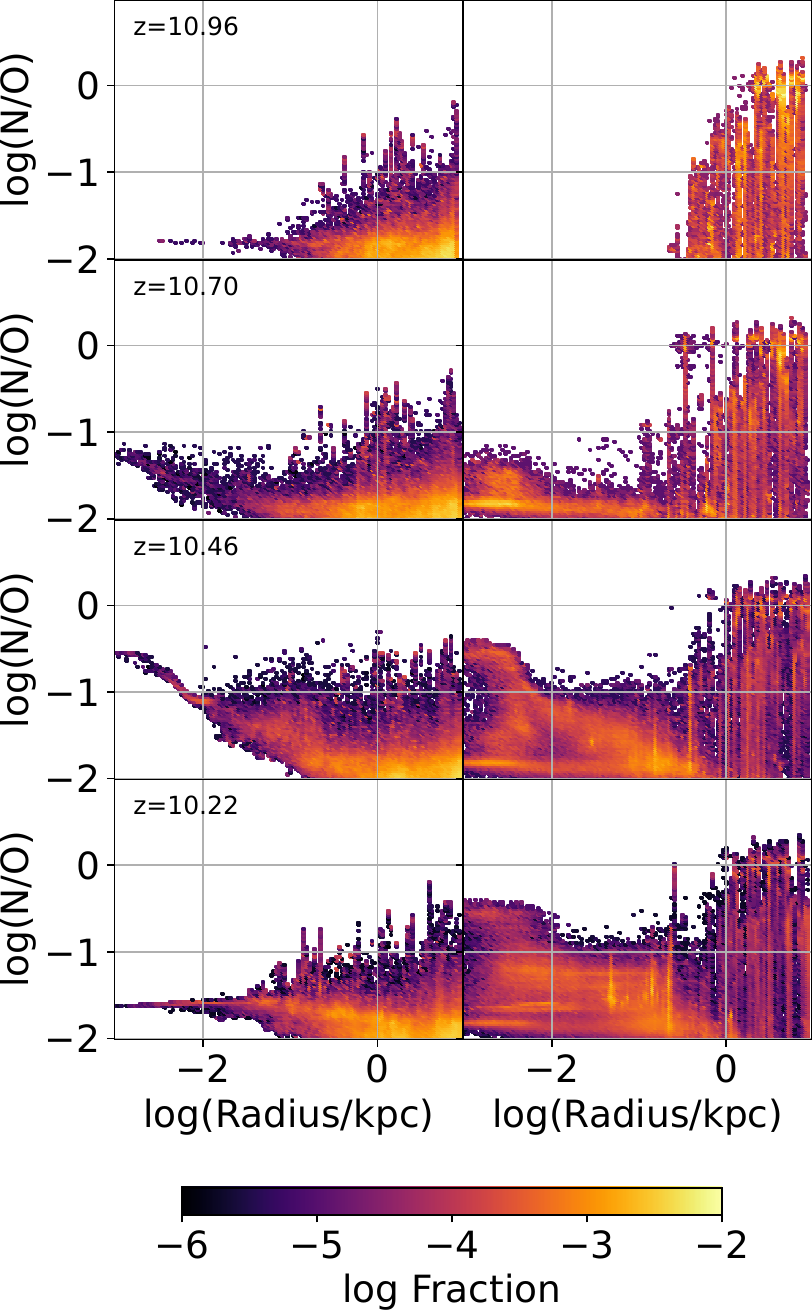}
 \end{center}
    \caption{Evolution of log(N/O) as a function of radius. The left column
    is the gas-phase metallicity while the right column is the stellar-phase
    metallicity. From top to bottom, we show the evolution of the ratio and the
    corresponding redshift is overplotted on the each panel. Color represents
    the relative mass fraction of grids in the log scale.
    {Alt text: Multi-panel density plot showing evolution of nitrogen-to-oxygen
    abundance ratio versus radius across four sequential epochs. Left column
    displays gas-phase ratio; right column shows stellar phase ratio. Color
    indicates relative mass fraction of grids in log scale.}
    }
\label{fig:phase:NO}
\end{figure}

The evolution of the gas-phase log(N/O) within the galaxy is shown in the left
column of figure~\ref{fig:phase:NO}. The central log(N/O) undergoes drastic
changes over a brief period. Initially, the log(N/O) is as low as that expected
by CCSNe. Subsequently, the central log(N/O) increases rapidly due to the
stellar winds. However, due to the low log(N/O) of accreted gas [see panel (c)
in figure~\ref{fig:phase:Rvel_OH12}], the wind yields are diluted and fail to
reach its idealized value. Eventually, the log(N/O) returns to the expected
value of CCSNe again, owing to the self-enrichment. The high log(N/O) period
lasts only $\sim 10~\mathrm{Myr}$ (see below for more details). In the outer
regions ($>1~\mathrm{kpc}$), there are spiky peaks which are the gas associate with
the star clusters (see section~\ref{sec:starclusters}).

The stellar-phase evolution is shown in the right column of figure~\ref{fig:phase:NO}.
Unlike the gas-phase, stellar-phase values are fully the integration
of the formation history since stars preserve the original composition of their
natal gas. The evolution of the central region reveals that star
formation begins in low log(N/O) gas, and then stars are formed from the gas
polluted by stellar winds. In contrast to the gas-phase distribution,
multiple values occur at a given radius within $100~\mathrm{pc}$. As before, the
spiky peaks found in the outer part are associated with the star clusters.

\begin{figure}[htb]
 \begin{center}
  \includegraphics[width=8cm]{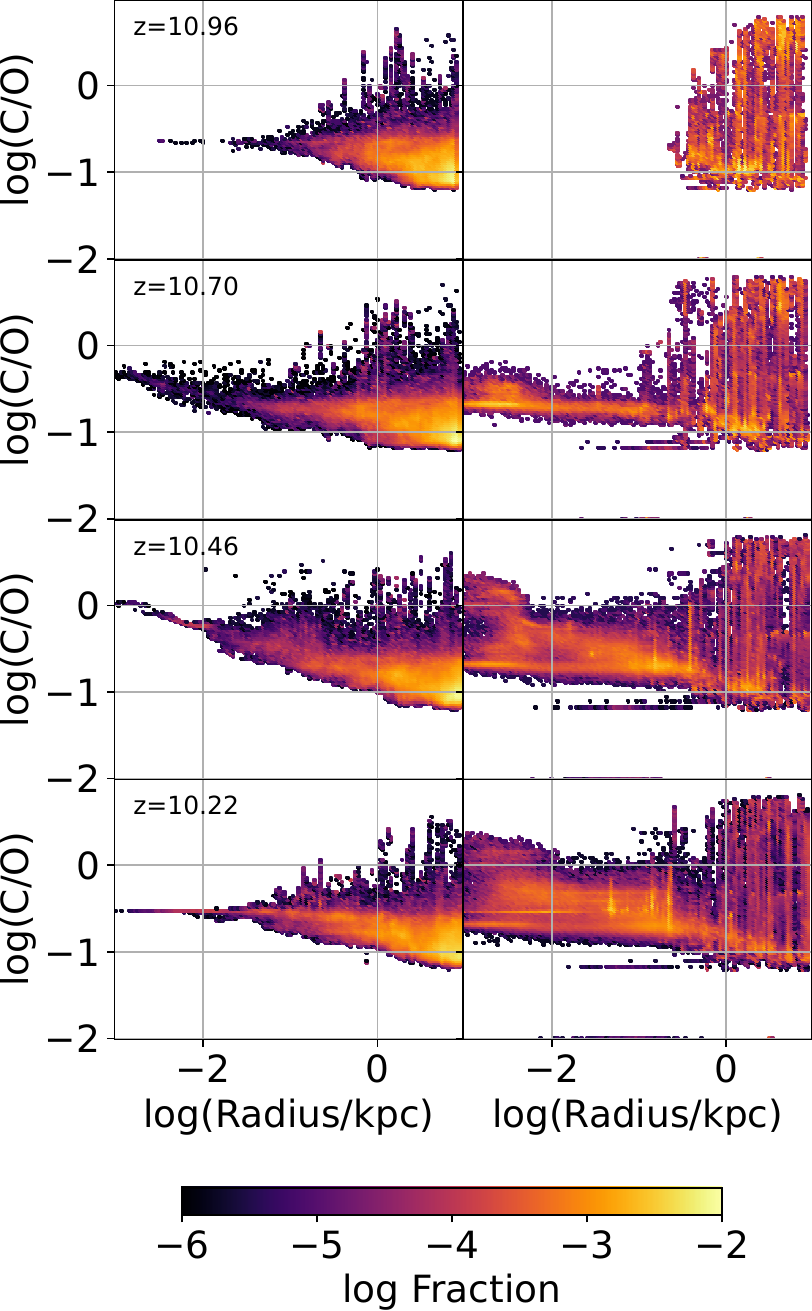}
 \end{center}
    \caption{Same as figure~\ref{fig:phase:NO}, but for log(C/O).
    {Alt text: Multi-panel density plot showing evolution of carbon-to-oxygen
    abundance ratio versus radius across four sequential epochs. Left column
    displays gas-phase ratio; right column shows stellar phase ratio. Color
    indicates relative mass fraction of grids in log scale.}}
\label{fig:phase:CO}
\end{figure}

We present the time evolution of the log(C/O) value in the galaxy in
figure~\ref{fig:phase:CO}. The overall evolution is the same as for log(N/O).
The log(C/O) value in the accretion gas, shown in panel (d) in
figure~\ref{fig:phase:Rvel_OH12}, is also very low.  Reflecting the fact that
the difference between the expected yields of CCSNe and the stellar winds is
rather small, the difference is also small compared to the case of log(N/O).

\begin{figure}[htb]
 \begin{center}
  \includegraphics[width=8cm]{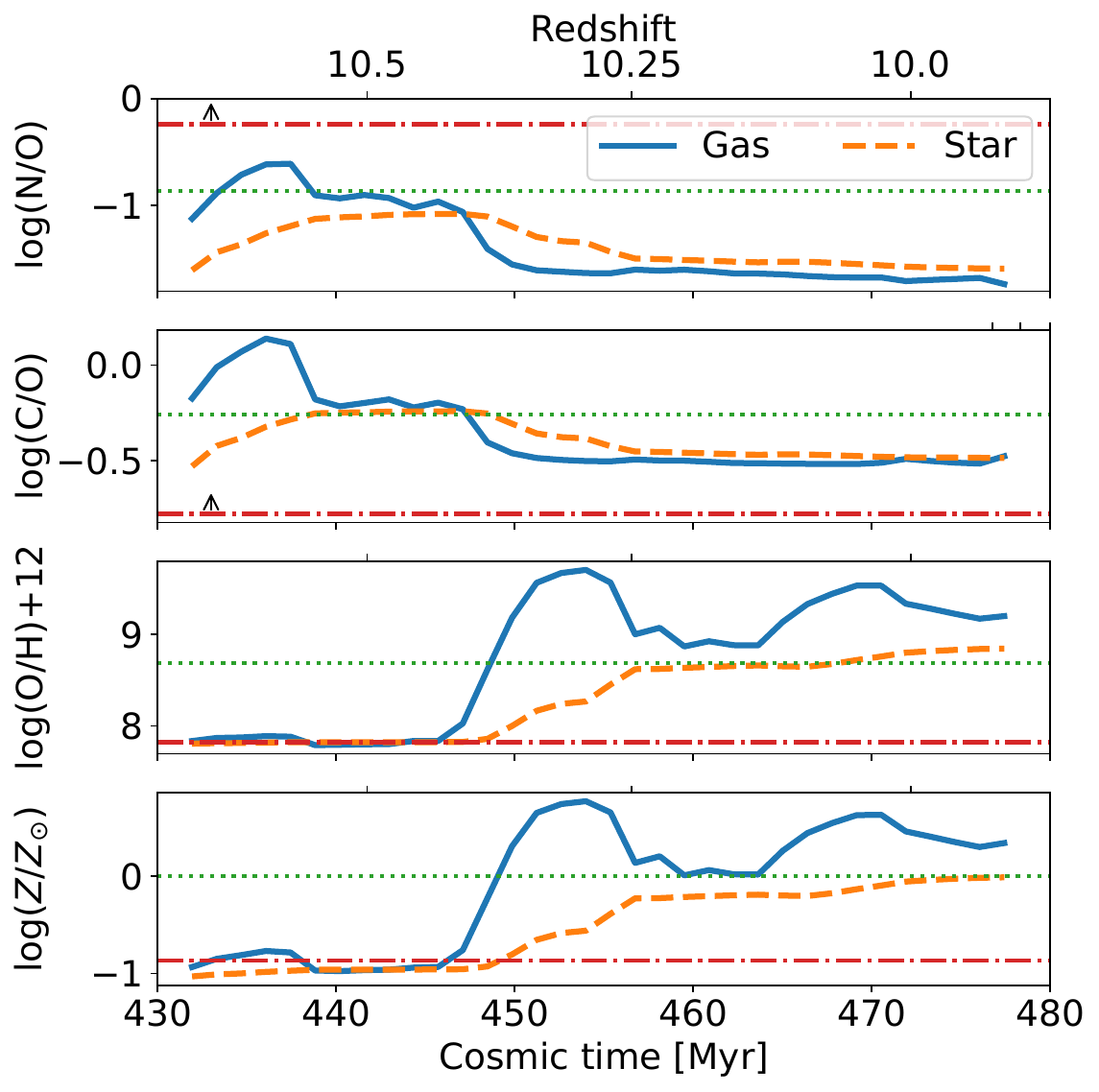}
 \end{center}
    \caption{The evolution of log(N/O), log(C/O), log(O/H)+12, and
    log($Z/\Zsun$) as a function of time and redshift, using high-frequency
    outputs during the central starburst. Blue solid curves represent the
    averaged values of the gas-phase in the central $10~\mathrm{pc}$ region,
    while orange dashed curves indicate those of the stars. Green dotted lines
    display the solar values, whereas red dot-dashed lines depict the (fiducial)
    values of GN-z11 \citep{Cameron+2023}. Arrows in the upper two panels
    indicate that these lines mean the minimum values.  Note that the
    metallicity of the GN-z11 is evaluated from the O/H ratio, while that of the
    simulation is directly counted from the elements.
    {Alt text: Multi-panel line plot showing time evolution of
    nitrogen-to-oxygen ratio, carbon-to-oxygen ratio, oxygen abundance, and
    metallicity during central starburst. Solid curves represent the central
    gas-phase values, while dashed curves represent the stellar-phase values.
    First two ratios increase at early phase around four hundred thirty megayear
    and gradually decrease, whereas the oxygen abundance and metallicity show
    sharp rises with about twenty megayear time lags.}
    }
\label{fig:evolution:chemistry}
\end{figure}

Figure~\ref{fig:evolution:chemistry} provides the more detailed time evolution
of the chemical compositions in the central region during the starburst phase.
We focus on the central $10~\mathrm{pc}$, in accordance with
figures~\ref{fig:phase:NO} and \ref{fig:phase:CO}. The baselines of log(N/O) and
log(C/O) in the pre-starburst phase, not shown in this figure, are the same as
in the post-starburst phase.

We observe the rapid evolution of the gas-phase log(N/O) and log(C/O), with
peaks at $t\sim436~\mathrm{Myr};z\sim10.60$. These peaks of log(N/O) and
log(C/O) are well above the solar abundance. Although the peak log (N/O) value
(-0.61) is lower than that of GN-z11, it is higher than that of JADES-GS-z9-0
and comparable to that of GN-z9p4. These values then decrease and form plateaus
near the solar values for $\sim 10~\mathrm{Myr}$. Finally, they become those
expected from the CCSNe yields (see figure~\ref{fig:YieldsComparison}).  The
strong synchronization among log(N/O), log(C/O), and SFR enhancements indicates
that stellar winds are the primary source of this pollution.

The peak values do not reach the pure stellar wind levels, because the gas
accretion onto the central region is already pre-polluted by CCSNe ejecta.
Radial profiles of the gas-phase reveal low log(N/O) and log(C/O) in the outer
regions (figures~\ref{fig:phase:NO} and \ref{fig:phase:CO}).  Efficient
mechanisms, such as a top-heavy IMF \citep{BekkiTsujimoto2023}, might be
necessary to obtain observational levels of log(N/O) under the stellar wind
pollution.

Stellar log(N/O) and log(C/O) values evolve more moderately.  Hence, there is a
non-negligible gap between the averaged gas-phase and averaged stellar-phase
metallicities in the early phase. The gap becomes almost negligible after
$t=460~\mathrm{Myr};z=10.20$. However, the succeeding starburst (refer to SFR in
figure~\ref{fig:SFR}) would reintroduce a short-lived gap.

In the third and fourth panels of figure~\ref{fig:evolution:chemistry}, we show
the evolution of the oxygen abundance and $Z$. A clear $\sim 20~\mathrm{Myr}$
lag reflects the CCSN time lag. In our model, we adopt the Limongi \& Chieffi
yields for non-Pop III stars. The massive stars that explode as CCSNe range from
$13~\Msun$ to $25~\Msun$. This results in the minimum and
maximum CCSNe time lags of $\sim 8~\mathrm{Myr} (25~\Msun)$ and
$\sim19~\mathrm{Myr} (13~\Msun)$ \citep{Portinari+1998, Saitoh2017}.
The event timing is computed based on the given IMF and thus the number of SNe
per unit time is biased toward longer lags; consequently the net enhancement
shows a characteristic lag of $\sim 20~\mathrm{Myr}$. The enhancement of the
gas-phase metallicity is much higher than that of the stellar-phase. This is
because the gas-phase reflects the instantaneous values, unlike the
stellar-phase.

The gas-phase values become the super-solar values while the stars
asymptotically approach the solar values. Numerical simulations, such as
FOREVER22 \citep{Yajima+2022, Yajima+2023}, predict these near-solar and
super-solar metallicities in high-$z$ galaxies (see figure~3 in
\cite{Isobe+2023}).

The overall evolution clearly shows that a high log(N/O) value is observed in
the galactic center at the very early stage of starburst.  Although it would
depend on the adopted yield table, the possible window is in the range of $\sim
10$--$20~\mathrm{Myr}$.

Note that this would be sensitive to the degree of metal enrichment in the
accreted gas. The model without Pop III mode (NoPopIII) does not show strong
concentration of star formation, reflecting enhanced metallicity and smaller
Jeans mass in the accreted gas. Star formation makes the accretion process less
dissipative. See section~\ref{sec:PopIII} for further details.

\begin{figure}[htb]
 \begin{center}
  \includegraphics[width=8cm]{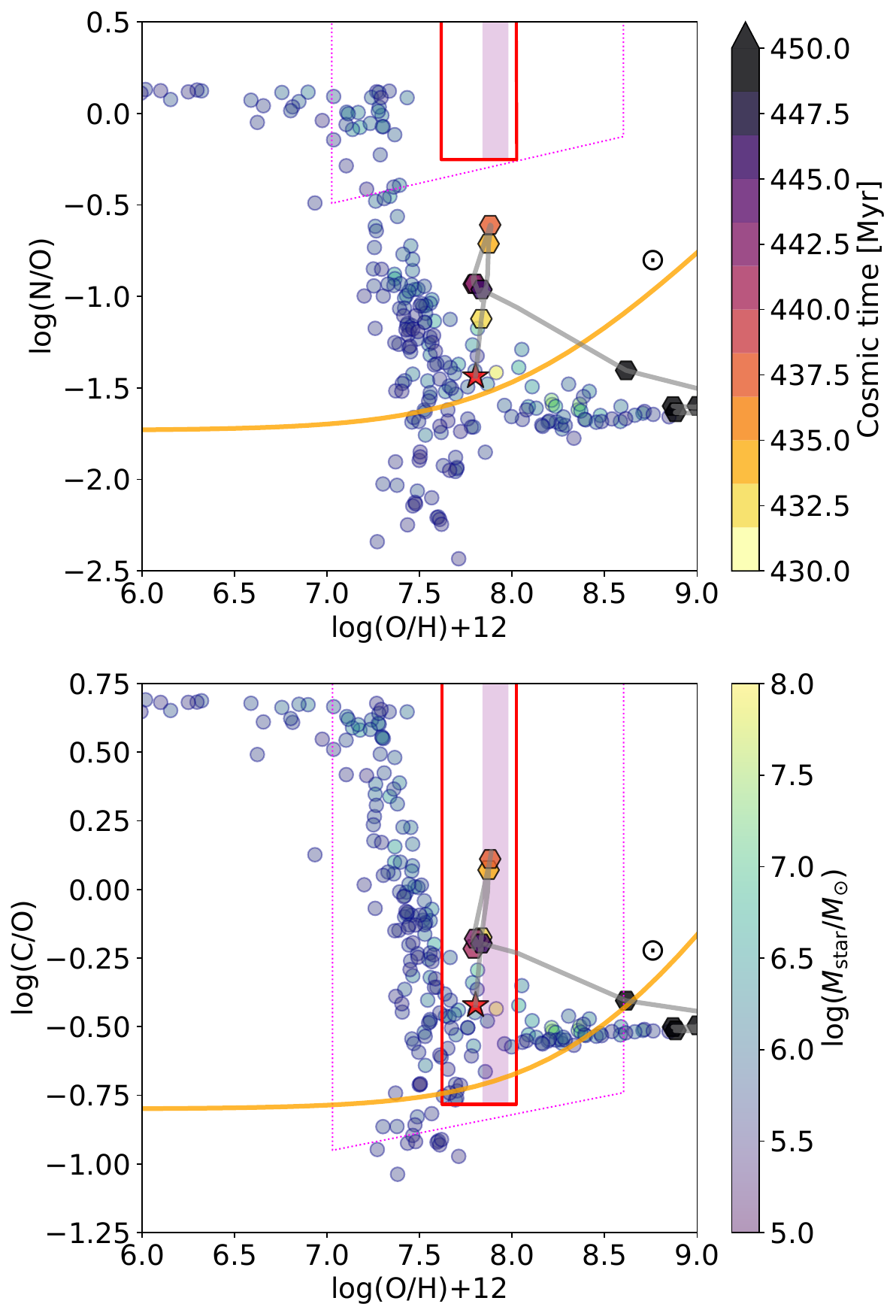}
 \end{center}
    \caption{Distributions of averaged log(N/O) and log(C/O) of the galactic
    center and star clusters in the galaxy as a function of log(O/H)+12.  The
    evolution tracks of the galactic center are shown with gray curves with
    symbols ($t=430$--$450~\mathrm{Myr};z\sim10.70$--$10.36$). Red stars represent the pre-starburst values
    ($t=431~\mathrm{Myr};z=10.70$).  We place hexagons every $2.76~\mathrm{Myr}$
    for reference. Colors represent the epoch and the corresponding color bar is
    shown at the right side of the top panel.  Each large circle, color-coded by
    its stellar mass, represents the values of star clusters in the primary
    halo. The color exhibits the stellar mass and the corresponding color bar is
    found at the right side of the bottom panel. For reference, we include the
    observational values (orange curves; \cite{Nicholls+2017}) and those of
    GN-z11 (magenta solid and dotted lines by \cite{Cameron+2023}; magenta
    filled region by \cite{Alvarez-Marquez+2025}). The solar symbols denote the
    solar abundance values \citep{Asplund+2009}.
    {Alt text: Two-panel scatter plot comparing nitrogen-to-oxygen and
    carbon-to-oxygen ratios against oxygen abundance. Gray curves with stars and
    hexagon symbols show evolutionary track of galactic center during starburst.
    Circles represent individual star clusters with stellar mass
    indicated for each cluster. Observational predictions for GN-z11 are also
    shown. Evolution of galactic center shows a significant and rapid
    enrichment in both abundance ratios during the period shown.}
    }
\label{fig:stat}
\end{figure}

The evolutionary trends of the central region on the log(N/O)-log(O/H)+12 and
log(C/O)-log(O/H)+12 planes are depicted in figure~\ref{fig:stat}. As shown in
figure~\ref{fig:phase:NO}, the log(N/O) value first increases due to stellar
winds keeping a low log(O/H)+12. After that, the CCSNe enhance oxygen.
Consequently, on the log(N/O)-log(O/H)+12 plane, the evolutionary track goes up
and then moves to the right as it goes down. If stellar wind alone is the only
contributor, the expected value is higher than the lower limit of GN-z11 (see
figure~\ref{fig:YieldsComparison}). However, the gas-phase mixing with the
preexisting oxygen prevents it from reaching the pure stellar wind value, as
evident in figure~\ref{fig:phase:NO}. The evolutionary track of log(C/O) is
similar to that of log(N/O). The log(C/O) falls within the range observed for
GN-z11.

\subsection{Star clusters at $z\sim10$ around the primary galaxy}
\label{sec:starclusters}

JWST has discovered very massive, compact star clusters at high-$z$ (e.g.,
\cite{Mowla+2022, Vanzella+2022, Vanzella+2023, Claeyssens+2023, Adamo+2024,
Fujimoto+2025, Messa+2024, Claeyssens+2025, Morishita+2025}). These clusters
might be precursors to present-day GCs. Our simulated galaxy also contains
numerous star clusters in its halo. Here, we focus on the star clusters found
around the primary galaxy at $z\sim10$.

In \citet{Dekel+2023}, they predict heavier Jeans mass, i.e., $\sim 10^{6}
~\Msun$ around the high-$z$ galaxies. With numerical simulations,
\citet{Kiyuna+2023} pointed out the Jeans mass of the post-shock gas becomes
$10^{4-5}~\Msun$ around the high-$z$ galaxies. This high Jeans mass
leads to the massive star cluster formation and/or supermassive star formation.
Since there is no specific recipe for the massive star formation in our
simulation, we focus on the star clusters.

We adopt the substructures identified by AHF to select star clusters, applying
the following criteria: (1) substructures without further substructures, and (2)
substructures whose stellar masses are required to exceed
$3\times10^5~\Msun$.  The second criterion yields star clusters with an
approximately simple power-law mass distribution and that is comparable with the
condition that $N_{\rm star} \ge 300$.

\subsubsection{Statistical distribution of star clusters}

\begin{figure*}[htb]
 \begin{center}
  \includegraphics[width=16cm]{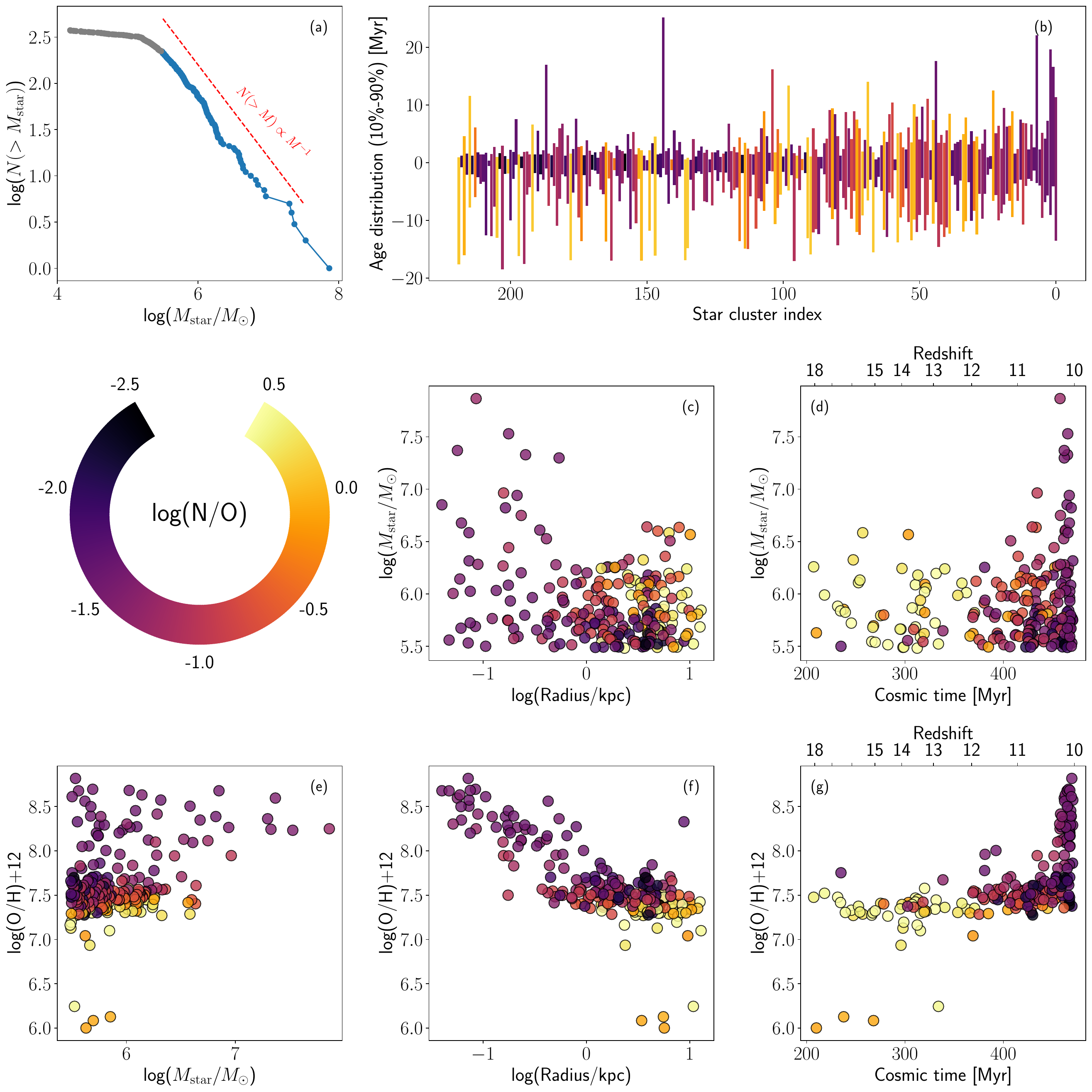}
 \end{center}
    \caption{Distributions of star clusters at $t=472~\mathrm{Myr};z=10.0$.
    Panel (a): cumulative mass function of star clusters. Blue points are
    the star clusters heavier than $3\times10^5~\Msun$, while the gray
    points are the rest. The red dashed line is a simple power-law function of an
    index of $-1$. For this study, we only analyze star clusters with masses
    exceeding $3\times10^5~\Msun$, as the break in the mass function
    implies poor detectability of lighter clusters due to limited mass
    resolutions.  Panel (b): age spread of the star clusters. We plot a line
    indicating the range from 10\% to 90\% of the ages, after setting the median
    age of each star cluster to be zero. The color indicates the log(N/O) value,
    with the color arc shown at the left-middle panel of this figure. This color
    code applies to all subsequent panels.  Panel (c): average metallicity of
    star clusters as a function of their stellar mass.  Panel (d): star cluster
    mass versus distance from the center of the host galaxy. Panel (e): star
    cluster metallicity versus distance from the center of the host galaxy.
    Panel (f): star cluster mass as a function of the mass-weighted age. Panel
    (g): star cluster metallicity as a function of mass-weighted age. Colors in
    panels (b)--(g) follow values of log(N/O) in individual star clusters, and
    the color arc is found in the middle-left of this figure.
    {Alt text: Multi-panel plot showing properties of star clusters. Panel (a)
    displays the cumulative mass function, which follows an inverse power law of
    negative one for clusters above three times ten to the power of five solar
    masses. Panel (b) shows the age spread of individual clusters. The remaining
    five scatter plots display relationships between cluster mass,
    metallicity, age, and distance from the galactic center.}
    }
\label{fig:starclusters:seven}
\end{figure*}

The cumulative mass function of the star clusters is presented in panel (a) of
figure~\ref{fig:starclusters:seven}. For masses above $M_\mathrm{star} =
3\times10^5~\Msun$, the mass function is closely approximated by a
single power-law function with an index of $-1$, as we have mentioned
previously.  A power-law function is widely confirmed for observed star cluster
systems (e.g., \cite{Whitmore+1999, Chandar+2010, Rodruck+2023}) and numerical
simulations (e.g., \cite{Saitoh+2011,Matsui+2019,ElmegreenLahen2024}).
{\footnote {It is also pointed out that observed star-cluster mass functions
have a cut-off, resulting in the Schechter function
\citep{PortegiesZwart+2010ARAA, Bastian+2012, Adamo+2015}. Numerical simulations
often show the Schechter function as star-cluster mass functions \citep{Li+2017}
. We do not observe any cut-off in our mass function. One possible reason is
that the mass function of our simulation is based on a very short window of the
extremely violent burst phase.  In fact, the numerical simulations of star
clusters that treat star clusters as collisional systems \citep{Grudic+2023,
Rodriguez+2023} have shown that the mass function of star clusters is
time-varying and that it is difficult to describe it by a single functional
form.
}}
A single power-law mass function implies that star cluster formation is primarily
driven by scale-free physics \citep{Elmegreen2009}. The low-mass part
($M_\mathrm{star} < 3\times10^5~\Msun$) is excluded in the current
analysis for possibly poor resolution.  The total star-cluster mass is
$4\times10^8~\Msun$, which is 18 percent of the total stellar mass of
the host galaxy at this time ($M_\mathrm{star} = 2.2\times10^9~\Msun$),
indicating a very high star cluster formation efficiency.

Figure~\ref{fig:stat} illustrates the chemical compositions of the star clusters
in the halo. We observe two distinct plateaus in the log(N/O) and log(C/O)
distributions, connected by narrow bridges around log(O/H)+12 = 7.5. The lower
plateau corresponds to the observations \citep{Nicholls+2017}, showing that it
results from CCSNe. Conversely, the higher plateau is attributed to
stellar winds alone.

The relatively short duration of star formation in individual clusters also
suggests that stellar winds and subsequent CCSNe are more likely to be the
primary polluters.  Panel (b) of figure~\ref{fig:starclusters:seven} depicts the
age distribution of the star clusters. In this panel, we set the median
formation time of each star cluster to zero, and plot a line indicating the
range from 10\% to 90\% of the formation time. The typical timescale for star
formation in individual star clusters is remarkably short, less than
$10~\mathrm{Myr}$. In the high-mass regime (star clusters with lower star
cluster indices), star clusters tend to have a longer formation timescale than
low-mass star clusters, but this trend is not prominent.  However, these are
still sufficiently short to exclude the contribution of AGB stars.  A similar
short formation timescale has been reported in numerical simulations of star
clusters at the similar mass scales \citep{Matsui+2019}.  This short timescale
of star formation is consistent with observations implying a short timescale for
star-cluster formation \citep{Bastian+2013}. A typical SFR in each star cluster
becomes $\sim 1$--$10~\Msun~\mathrm{yr}^{-1}$, which is a natural
consequence of the short timescale.

It is worth noting that we exclude one star cluster from the analysis shown in
panel (b) of figure~\ref{fig:starclusters:seven}, as its formation timescale
exceeds $100~{\mathrm{Myr}}$. Our analysis reveals that this outlier results
from cluster mergers. Its age distribution consists of three distinct age groups,
each with a spread of $\sim 10~\mathrm{Myr}$. This pattern indicates that each
of the sub-clumps underwent a formation process similar to the others. This can
make distinct populations in a cluster, resulting in multiple stellar
populations (e.g., \cite{BastianLardo2018,Gratton+2019,Marino2020,
MiloneMarino2022}).

The mass distribution of star clusters as a function of radius, panel (c) in
figure~\ref{fig:starclusters:seven}, reveals that massive star clusters tend to
be located in the inner region and have lower log(N/O) values. On the other
hand, we can observe that in the outer regions ($R > 1~\mathrm{kpc}$), there is a
mix of star clusters both high and low log(N/O) values. Particularly, star
clusters around $R\sim 10~\mathrm{kpc}$, which is comparable with the virial radius
at this epoch, have higher log(N/O) values. These star clusters are likely to be
accreted clusters.

The mass distribution of star clusters as a function of formation time [panel (d)
in figure~\ref{fig:starclusters:seven}] offers additional insight: there is a
clear tendency for older star clusters to have higher log(N/O) values. With
these findings, we conclude that star clusters with higher log(N/O) are old,
accreted clusters. They likely formed in less massive minihalos, i.e., ex-situ
origins. In contrast, the high-mass star clusters identified in this simulation
are likely in-situ origins that originated during the starburst. These star
clusters correspond to the prediction of the feedback-free starburst model
\citep{Dekel+2023}.

The metallicity distributions as functions of mass, radius, and formation time
are exhibited in panels (e), (f), and (g) of figure~\ref{fig:starclusters:seven}.
These panels reveal that massive star clusters born in the inner region
during the starburst tend to have higher metallicities [log(O/H)+12$>7.7$;$>
0.1~\Zsun$].  These clusters do not have higher log(N/O) values due to the
difficulty of diluting the oxygen present in the natal gas.

Recent observations of the star-forming clumps at $z=7.88$ by
\cite{Morishita+2025} reveal a significant metallicity scatter among them. These
clumps have masses ranging from $10^8$ to $10^9~\Msun$, with metallicities
[log(O/H)+12] spanning from $<7$ to $\sim8$. As shown in panel (e)
of figure~\ref{fig:starclusters:seven}, our simulated star clusters exhibit
similar metallicity diversity. In the high-mass range ($>10^6~\Msun$),
the star clusters' metallicities span nearly 1 dex [log(O/H)+12=7.5--8.5],
while there is an even greater spread in the low-mass range
($<10^6~\Msun$). Although our simulations show a slightly higher
averaged metallicity ($\sim0.5$~dex), the metallicity span is consistent with
their observations.

One more interesting finding is the presence of four extremely metal-poor star
clusters [log(O/H)+12 $\sim6$] with masses below $10^6~\Msun$ [panels
(e), (f), and (g) in figure~\ref{fig:starclusters:seven}]. These clusters have
ex-situ origins, likely forming in less massive minihalos filled with pristine
gas. \citet{Vanzella+2023} have discovered Pop III star cluster candidates at
$z=6.64$ by gravitational lensing. While their mass range is one to two orders
of magnitude smaller than ours, their metallicity range matches ours. These
observed clusters, LAP1 and LAP1-B, may have formed in minihalos.

\subsubsection{Stellar populations in individual star clusters}

We investigate the chemical compositions of star particles in star clusters.
Here, we examine the twenty-four representative star clusters located more than
$5~\mathrm{kpc}$ from the galaxy center.  This is because inner clusters are
expected to sink toward the galaxy center due to the dynamical friction, giving
them little chance of surviving low redshift.

\begin{figure*}[htb]
 \begin{center}
  \includegraphics[width=16cm]{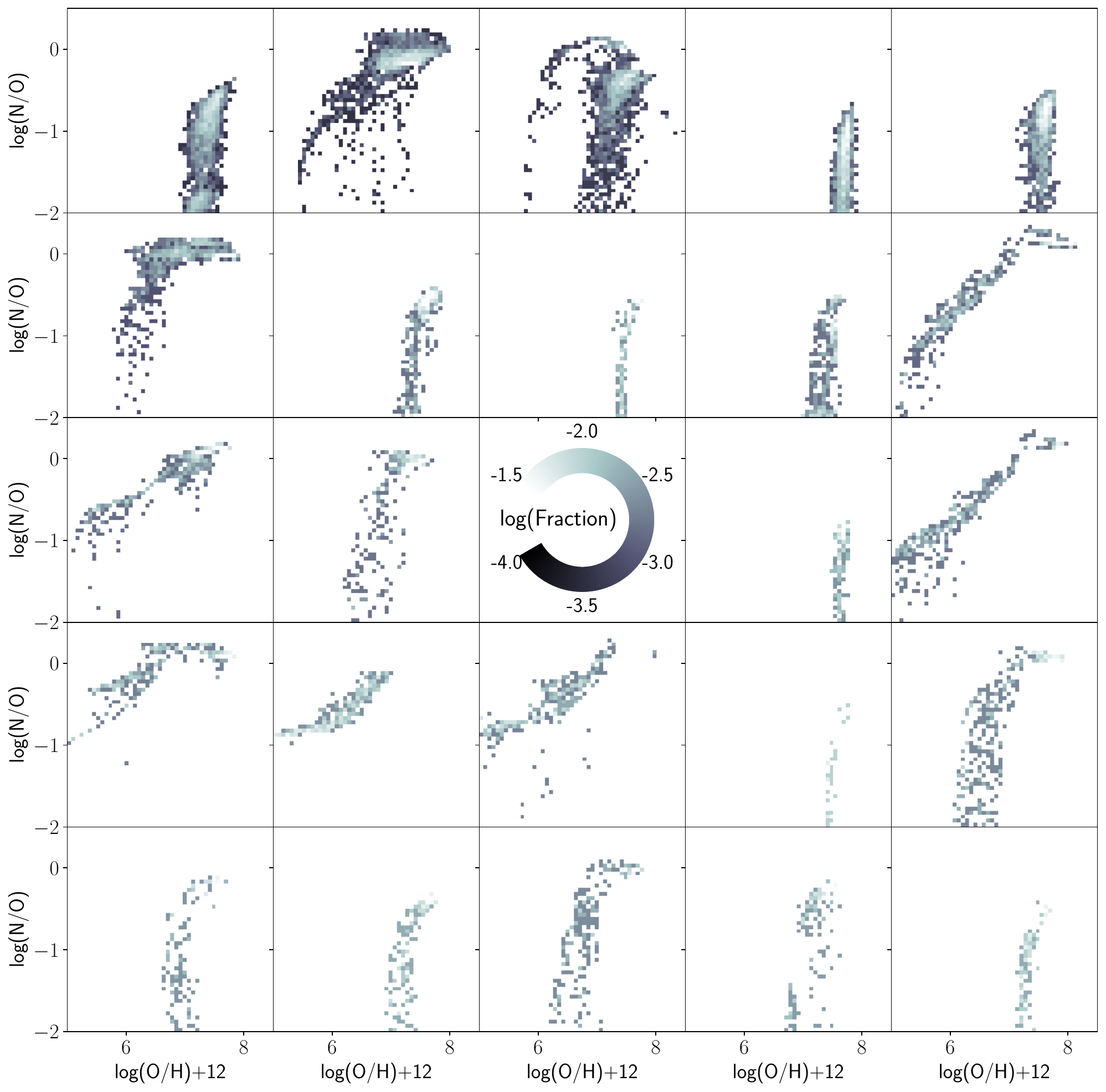}
 \end{center}
    \caption{Distributions of star particles on the metallicity-log(N/O) plane.
    The snapshot at $t=472~\mathrm{Myr};z=10.00$ is adopted. Color follows the
    fraction of star particles in each grid.
    {Alt text: Multi-panel plot with five rows and five columns showing
    distributions of star particles in the nitrogen-to-oxygen versus oxygen
    abundance plane. Individual panels display metallicity distributions of star
    particles for each star cluster. Central panel shows reference scale for
    particle fraction in each grid.}
    }
\label{fig:starclusters_NO}
\end{figure*}

We can investigate their formation histories by looking at the distributions of
log(N/O) values as a function of metallicity. In figure~\ref{fig:starclusters_NO},
we show the distributions of star particles on the metallicity-log(N/O) plane.
While there are variations, we can summarize their evolution as starting from
low-metallicity, low-log(N/O) state and increasing to a high-metallicity,
high-log(N/O) state. We note that most stars are concentrated in the
high-log(N/O) state. This indicates that the evolution timescale is very short
(see panel (b) of figure~\ref{fig:starclusters:seven}) and the evolution
suddenly finishes via feedback, i.e., CCSNe.

Although all of them evolve from left-bottom to right-top, two subgroups emerge
in their evolution patterns. One shows a shallow gradient, and the other
exhibits a steep gradient. These trends correlate with the initial
metallicities. Stellar wind, the primary polluter, releases only a small amount
of oxygen, resulting in small horizontal evolution. When the original gas
contains sufficient amount of oxygen, horizontal evolution becomes hard to see.
The steep gradient group surely initiates from relatively metal-rich gas, hence
they have such a gradient. On average, the gas-phase metallicity increases with
time. The analysis of the formation time reveals that the star clusters with
steep gradients form at the late stage of the simulation, i.e., $t>
400~\mathrm{Myr};z<11$.

One cluster displays an interesting evolution track. In the top-center panel, we
observe two distinct tracks. The dominant one originates from gas with log(O/H)
+12 $\sim7$, while the other begins in that of log(O/H)+12 $\sim6$. The cluster
comprises (at least) two distinct stellar populations originate from different
gas clouds. As expected, there is a gap of $\sim 5~\mathrm{Myr}$ between the
formation times of these two components.

It is obvious that stars are likely to have high log(N/O) values and the most
frequent value of log(N/O) is around zero. This high log(N/O) value is
comparable to the local member stars in GCs (e.g., \cite{BastianLardo2018,
Gratton+2019, MiloneMarino2022}). Through chemodynamical analysis of APOGEE and
Gaia data, \citet{BelokurovKravtsov2023} show that stars with high log(N/O)
values comparable to GN-z11 originate from the massive star clusters.  According
to our simulation, high log(N/O) stars (i.e., log(N/O) $\sim0$) are found only
in the star clusters, their members would eventually be spread throughout the
host halos at low-$z$.

Observations show that GCs have anomalous metallicity variations among their
member stars, unlike the conventional view. A particularly notable feature is
the anti-correlation between the oxygen and sodium (Na) abundances, known as the
Na-O anti-correlation (e.g., \cite{Carretta+2006, YongGrundahl2008,
BastianLardo2018, Carretta2019}).  Various explanations are proposed for this
phenomenon, such as AGB pollution (e.g., \cite{D'Ercole+2008}), fast-rotating
massive stars (e.g., \cite{Decressin+2007}), very massive stars (e.g., \cite
{DenissenkovHartwick2014}), binary interactions (e.g., \cite{deMink+2009}),
mergers of massive stars \citep{Wang+2020MSP}. Despite these efforts, no
consensus has been reached on the origin of this feature \citep{Renzini+2015,
BastianLardo2018}.

\begin{figure*}[htb]
 \begin{center}
  \includegraphics[width=16cm]{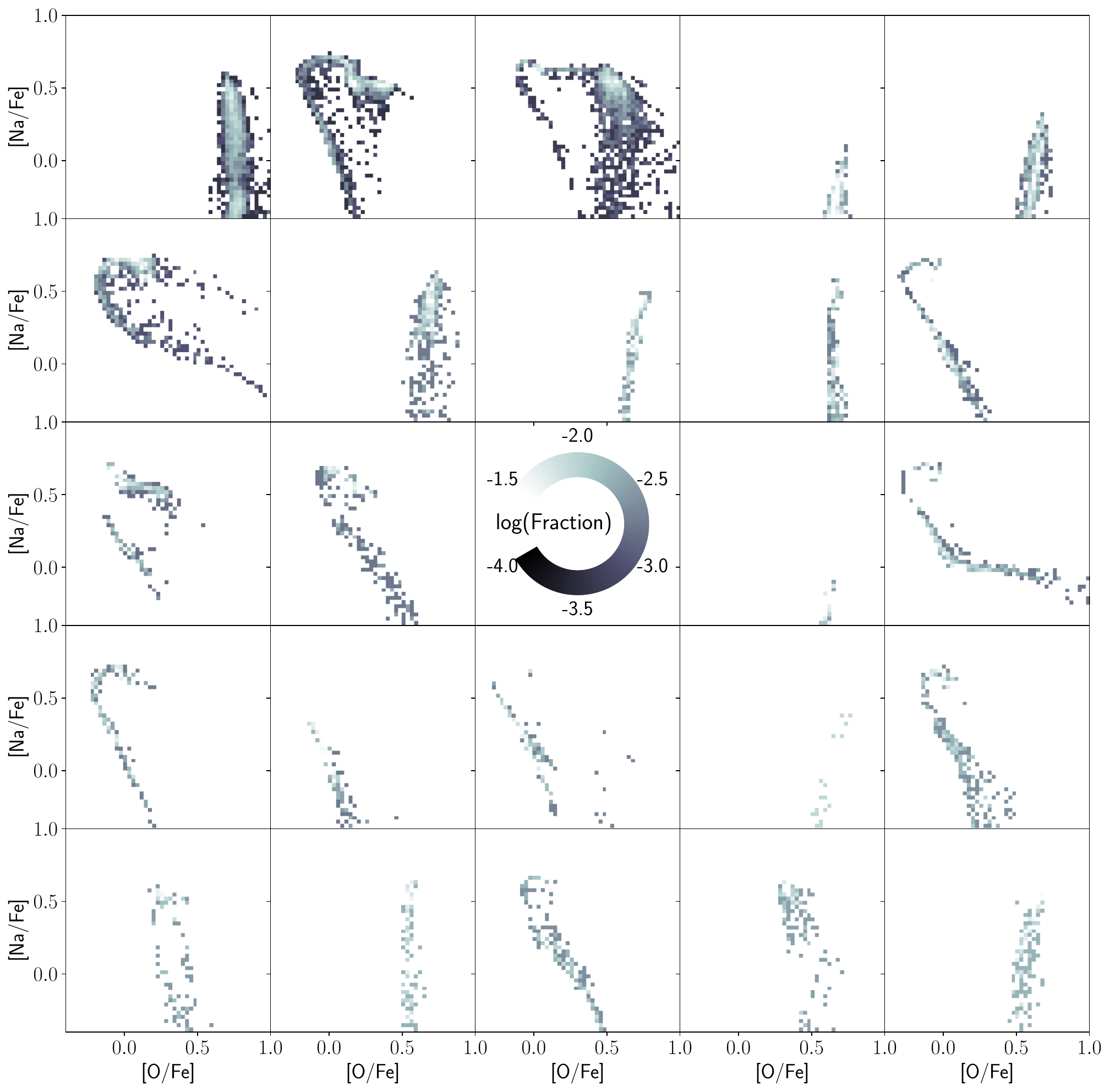}
 \end{center}
    \caption{Same as figure~\ref{fig:starclusters_NO}, but on the [Na/Fe]-[O/Fe] plane.
    {Alt text: Multi-panel plot with five rows and five columns showing
    distributions of star particles in the sodium-to-iron ratio versus
    oxygen-to-iron ratio plane. Individual panels display metallicity
    distributions of star particles for each star cluster. Central panel shows
    reference scale for particle fraction in each grid.}
    }
\label{fig:starclusters_NaFe}
\end{figure*}

\begin{figure}[htb]
 \begin{center}
  \includegraphics[width=8cm]{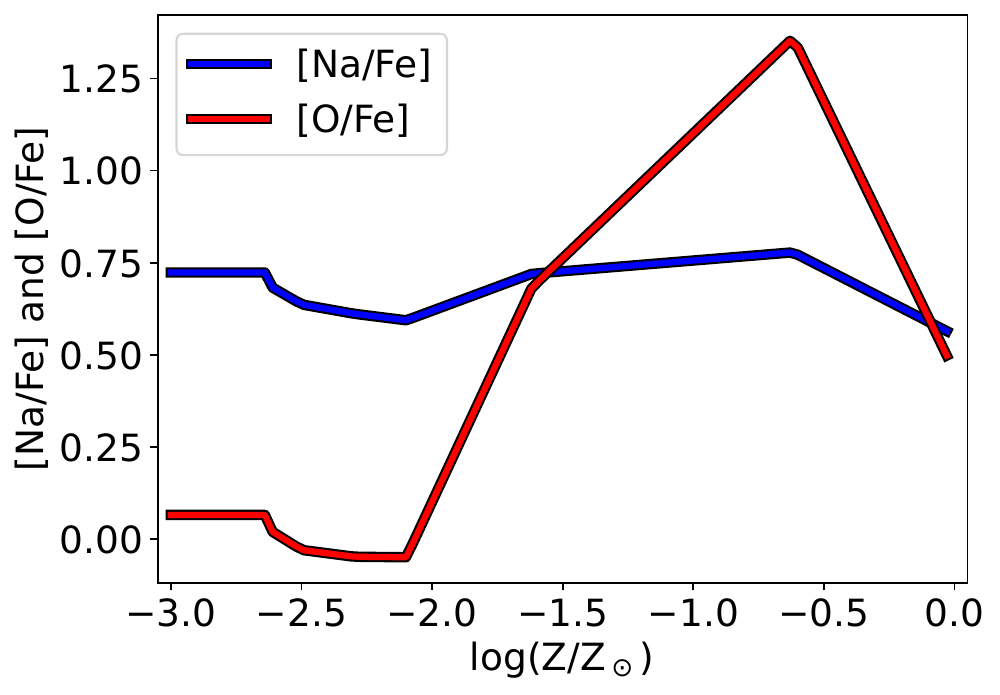}
 \end{center}
    \caption{IMF-weighted released amount of sodium and oxygen normalized by
    iron (Fe) as a function of metallicity. Only the wind composition is
    considered. The definition of the vertical axis is
    $[\mathrm{X}/\mathrm{Fe}] \equiv \log_{10}(n_\mathrm{X}
    /n_\mathrm{Fe})-\log_{10}(n_\mathrm{X}/n_\mathrm{Fe})_{\odot}$.
    Solar abundances follow \citet{Asplund+2009}.  The distribution of stellar
    rotation velocities \citep{Prantzos+2018} is taken into account.
    {Alt text: Line plot displaying initial mass function weighted
    sodium-to-iron ratio and oxygen-to-iron ratio versus metallicity for stellar
    winds. Sodium-to-iron ratio exceeds oxygen-to-iron ratio at lower
    metallicities, while oxygen-to-iron ratio dominates at higher metallicities.
    Near solar metallicity, sodium-to-iron ratio is slightly higher than
    oxygen-to-iron, but the differences are small.}
    }
\label{fig:NaFeOFe}
\end{figure}

We show the relations of [Na/Fe]-[O/Fe] for twenty-four representative star
clusters in figure~\ref{fig:starclusters_NaFe}. Over half of these clusters
exhibit Na-O anti-correlations among their member stars. As shown in
figure~\ref{fig:NaFeOFe}, in our model, stellar winds with very low metallicity
($Z\le10^{-1.5}~\Zsun$) release Na-rich but O-poor ejecta. Consequently, only
star clusters formed from low-metallicity gas display Na-O anti-correlations.
Although AGB stars are occasionally argued as a possible origin of this
anti-correlation, their contribution is unlikely due to the brief formation
timescale of these clusters.

\citet{Lahen+2024} argue for the establishment of the Na-O anti-correlation
through stellar winds. Their high-resolution numerical simulations of
galaxy-galaxy mergers tracked star cluster formation with considering chemical
compositions. Despite targeting a different redshift and physical conditions, our
simulation yields similar results. Both studies emphasize the crucial role of
initial metallicity in this process.

\section{Model Comparisons} \label{sec:comparisons}

Here, we compare the fiducial model and others. The contributions of Pop III
stars (model NoPopIII) are discussed in section~\ref{sec:PopIII}. The
contributions of the LWB strength (fiducial and NoLWB) are studied in
section~\ref{sec:LWB}.  Contributions of a top-heavy IMF, AGB stars, and
different AGB yield tables are investigated in sections~\ref{sec:TopHeavy} and
\ref{sec:AGB}.

\subsection{Contributions of Pop III stars} \label{sec:PopIII}

Pop III stars have large impacts on the early phase of galaxy formation due to
their expected top-heavy IMFs (e.g., \cite{Abel+2002, Bromm+2002,
Yoshida+2008, Susa+2014, Hirano+2015,Chon+2024}).  The increase in the relative
fraction of high-mass stars indicates that the released energy is larger, and
thus Pop III stars are more likely to destroy their host minihalos, resulting in
the large-scale metal ejection and causing the substantial enrichment of the
inter galactic medium (e.g., \cite{WiseAbel2008III,Mead+2025}).

\begin{figure}[htb]
 \begin{center}
  \includegraphics[width=8cm]{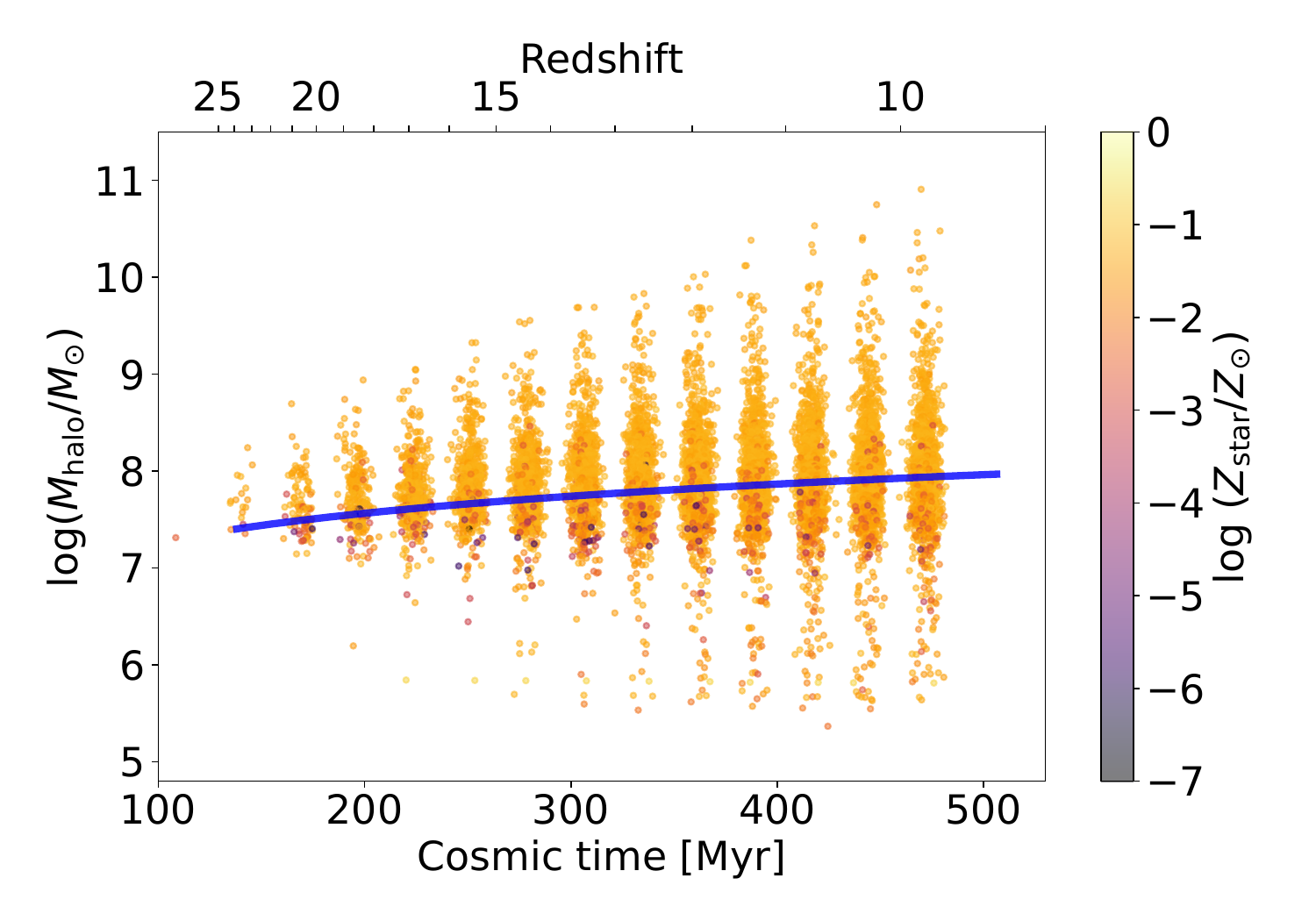}
 \end{center}
    \caption{Same as figure~\ref{fig:HaloMassStellarZ}, but for the case without
    the Pop III stars.
    {Alt text: Scatter plot showing halo mass versus time and redshift. Dots
    represent individual halos, color-coded by stellar metallicity. A reference
    curve shows expected halo mass for virial temperature of ten to the fourth
    Kelvin.}
    }
\label{fig:HaloMassStellarZnoPopIII}
\end{figure}

The model where we turned off Pop III star formation (i.e., all star particles
follow the Chabrier IMF) shows strong self-enrichment in all of the halos.  The
distribution of star-forming halo masses as a function of time (redshift) for no
Pop III model is shown in figure~\ref{fig:HaloMassStellarZnoPopIII}. Unlike the
case with the Pop III model (figure~\ref{fig:HaloMassStellarZ}), the averaged
metallicity of stellar components in each star-forming halo becomes very high.
This high average metallicity would lead to different evolution for galaxy
formation and also offsets in the chemical evolution in galaxies and star
clusters.

One crucial difference lies in the accumulation process of early galaxies.  Once
the accretion gas becomes metal-rich, star formation takes place within the gas
during the accretion phase due to its small Jeans mass thanks to efficient
cooling. This changes the mass accretion onto the galactic center to be less
dissipative. Consequently, there is less concentration, which would conflict
with the observations.

While the very early phase of the chemical evolution is completely different,
there is no large offset in the averaged metallicity of the most massive galaxy
comparing with the fiducial model at the late stage.
As far as our numerical experiments, strong feedback from Pop III stars is
important to explain ``compactness'' of high-$z$ galaxies.

\subsection{Contributions of the intensity of the Lyman-Werner background}
\label{sec:LWB}

The LW radiation plays a crucial role in the early phase of galaxy formation
(e.g., \cite{Ricotti+2001,Abel+2000,GalliPalla2013, Safranek-Shrader+2012,
Safranek-Shrader+2014,Nebrin+2023}). The LW radiation dissociates H$_2$
molecules through the Solomon process \citep{Solomon1965}. This process leads
to the suppression of star formation in the gas leading to a highly concentrated
star formation in the galactic centers of high-redshift galaxies, because of
dissipative effects.

When we turned off the LWB in our model (model NoLWB), early star formation was
enhanced and occurred in less concentrated regions before reaching the galactic
center. As a result, the star formation peak decreased by nearly one order of
magnitude, and we no longer observed a high concentration of star formation.
This comparison highlights the crucial role of LWB in generating the extremely
star-forming galaxies observed at high-$z$.

\subsection{Effects of top-heavy IMF for Pop II/I stars} \label{sec:TopHeavy}

In this subsection, we examine the contributions of a different IMF. We used the
snapshot of our fiducial model at $t=431~\mathrm{Myr};z=10.70$ and restarted
simulations with modified parameters.

Our fiducial model adopts a top-heavy IMF only for Pop III and the Chabrier IMF
otherwise. The transition metallicity between the Pop III IMF and the normal
IMF is set at $Z_\mathrm{PopIII} = 10^{-5} \Zsun$.  Above this threshold, stars
form under the Chabrier IMF. However, this modeling oversimplifies star
formation. Even in intermediate-metallicity gas, the IMF might be top-heavy
\citep{Chon+2024}, albeit the change would be more moderate compared with the
Pop III case. Observations suggest the existence of top-heavy IMFs even in the
local Universe (e.g., \cite{Stolte+2005, Schneider+2018}). Slopes of IMFs are
likely related to local environments, i.e., (surface) star formation rate
(e.g., \cite{Gunawardhana+2011,Salvador+2025}).  For massive galaxies, usually
bottom heavy IMFs are favorable in the local Universe, while top-heavy ones
would be necessary to explain a part of the excess of UV light from high-$z$
ones \citep{vanDokkumConroy2024, Jeong+2025}.

Figure~\ref{fig:evolution:chemistry} shows the central metallicity is almost 0.1
$\Zsun$, while the rapid gas accretion induces strong collisions in the galaxy
center. This prompts us to consider the effects of adopting different IMFs.

We modified the IMF form from the Chabrier IMF to a top-heavy one, using a
single power-law form with an index of $-0.35$. This change increases the
fraction of massive stars (those exceeding $10~\Msun$) by three times. This IMF
is also used in \citet{BekkiTsujimoto2023}. This is the TopHeavy model.

While the overall evolution of the galaxy remains comparable to the fiducial
model, we observe an increase in the log(N/O) value by 0.25 dex. Our fiducial
model already shows a significant increase of log(N/O) in the central region,
though it is less than the observational prediction by \citet{Cameron+2023}. The
0.25 dex increase brings our results to lower limit of the observational
prediction. This increase is slightly higher than that reported by
\citet{BekkiTsujimoto2023}. One of the clear reason is that our comparison is
the Chabrier IMF case and the top-heavy one. The other possible reason is, in
our simulations, the log(N/O) increase is highly localized to the central
region. This locality of pollution is a crucial factor not captured by one-zone
models.

\subsection{Contributions of AGBs} \label{sec:AGB}

While the timescale is much longer than that of stellar winds, AGB stars also
release nitrogen-rich and oxygen-poor gas.  A temporary suspension of star
formation is required to build an AGB ejecta-rich environment while avoiding the
contribution of CCSNe enrichment.  In our model, there are no major bursts
preceding the primary one at $z=11-10$, implying a limited AGB contribution.

To confirm this, we artificially deactivated the AGB mass return (gas and metals)
from $t=431~\mathrm{Myr};z=10.70$ and evolved this model for $100~\mathrm{Myr}$
(model NoAGB). This modification results in a slight (less than 0.1 dex)
decrease in the log(N/O) value at the galactic center. As expected, in galaxies
with a nearly monotonic SFR evolution, the contribution of AGBs is very limited.

We next assess the impact of different AGB yield sets (FRUITY vs Monash).  As
shown in figure~\ref{fig:YieldsComparison}, the Monash yield tables predict a
higher log(N/O) [a lower log(C/O)] value than the FRUITY yield tables,
reflecting the different efficiencies of the CNO cycle owing to higher
temperatures at the base of the convective envelope, which enhances the CNO
cycle. The Monash run, which also lacks a large break of star formation, showed
no clear log(N/O) enhancement, similar to the fiducial FRUITY run. These results
indicate that the star formation history strongly affects the enrichment of the
galactic center at high-$z$. The higher N production efficiency of the CNO cycle
in the Monash yields does not manifest on such short timescales.

Recently, \citet{McClymont+2025} presented the chemical evolution results from a
high-resolution suite of cosmological zoom-in simulations.  In their simulations,
N-enhanced environments appear due to the AGB pollution during the suppression
of star formation immediately following a starburst.  Key characteristics
include an N-enhanced environment that emerges after a starburst, with a
duration of $\sim 100~{\rm Myr}$. A notable feature is a clear offset from the
star-forming main sequence (SFR-$M_\mathrm{star}$), caused by the halt of star
formation.

In contrast, our simulations show an N-enhanced environment during the bursty
star formation because the primary polluter is the stellar wind from the massive,
fast-rotating stars, not AGBs. Consequently, the duration is much shorter than
that of the AGB scenario. The N-enhanced state is only visible in the central
region ($R\lesssim10~\mathrm{pc}$) on a timescale of $\sim 10~\mathrm{Myr}$,
before the contribution of CCSNe becomes significant. This stellar wind
scenario does not predict a clear offset from the star-forming main sequence
(see figure~\ref{fig:MstarSFR}). Reflecting the efficiency of N synthesis, the
log(N/O) value is slightly lower than the AGB model
(see figure~\ref{fig:YieldsComparison}), although this depends strongly on the
adopted yields.

\citet{McClymont+2025} argue that their AGB scenario predicts a lower $\mathrm{H}
\beta$ equivalent widths, $\mathrm{EW}(\mathrm{H}\beta)$, where most cases are
$\mathrm{EW}(\mathrm{H}\beta) < 40$~\AA\ and only rare cases exceed
$100$~\AA. In contrast, some N-enhanced galaxies show higher $\mathrm{EW}
(\mathrm{H}\beta)$ values, e.g., $> 100$~\AA. This can be understood
because $\mathrm{EW}(\mathrm{H}\beta)$ is sensitive to recent star formation. In
the AGB scenario, the longer duration of $\sim 100~\mathrm{Myr}$ makes it
difficult to maintain higher values during the N-enhanced state.  Therefore,
they argue that observed N-enhanced galaxies correspond to a rejuvenation phase,
during which galactic-scale gas with average abundances accretes onto the
N-enhanced clouds polluted by AGB stars.

Although we do not directly compute $\mathrm{EW}(\mathrm{H}\beta)$ from our
simulations, the bursty star formation that accompanies the N-enhanced state
would naturally yield large $\mathrm{EW}(\mathrm{H}\beta)$. \citet{Inoue2011}
provides the expected $\mathrm{EW}(\mathrm{H}\beta)$ for high-$z$ galaxies as a
function of age and metallicity.  Assuming a metallicity of $\sim 0.1 \Zsun$, an
average stellar age of $4\times10^6~\mathrm{yr}$ at $t\sim436~\mathrm{Myr};
z\sim10.60$, and the observationally predicted low escape fraction of $\sim
0.033$ \citep{Bunker+2023}, the expected $\mathrm{EW}(\mathrm{H}\beta)$ for our
galaxy is $\sim 350$~\AA. As an independent check, we can also infer
the $\mathrm{EW}(\mathrm{H}\beta)$ using an empirical approach based on
observations. To infer the possible value of $\mathrm{EW}(\mathrm{H}\beta)$, we
use the anti-correlation between $\mathrm{EW} (\mathrm{H}\beta+[\mathrm{O\,
III}])$ and the average stellar age at $z=5.5$--$7$ in \citet{Caputi+2024} and
adopt the typical EW ratio of $\mathrm{EW}(\mathrm{H}\beta) / \mathrm{EW}
([\mathrm{O\,III}]) \sim 0.1$.
{\footnote {\citet{Caputi+2024} show that $\log(\mathrm{EW}(\mathrm{H}
\beta+[\mathrm{O\, III}])/{\mbox{\AA}}) \sim -0.36 \log(\mathrm{Age}
/\mathrm{yr}) + 5.4$ for galaxies at $z = 5.5$--$7$. Our simulations predict
$\mathrm{EW}(\mathrm{H}\beta+[\mathrm{O\,III}]) \sim 1000$~\AA\ with an
average stellar age of $\sim 4\times10^{6}~\mathrm{yr}$ within the central
$10~\mathrm{pc}$ at the peak of log(N/O). We can convert this into $\mathrm{EW}
(\mathrm{H}\beta)$ if we have $\mathrm{EW}(\mathrm{H}\beta)/\mathrm{EW}
([\mathrm{O\,III}])$.  According to \citet{Boyett+2022}, based on the
observations of \citet{Tang+2019}, $\mathrm{EW}(\mathrm{H}\beta) = 0.115 \times
\mathrm{EW}([\mathrm{O\,III}] \lambda5007\,\mbox{\AA})^{1.065}$ and
$\mathrm{EW}(\mathrm{H}\beta+ [\mathrm{O\,III}]) = 1.48 \times \mathrm{EW}
([\mathrm{O\,III}]\lambda5007\,\mbox{\AA})$.  Thus, $\mathrm{EW}(\mathrm{H}
\beta) /\mathrm{EW}(\mathrm{H}\beta+[\mathrm{O\,III}]) \sim 0.1$. We note that
the relations used here were obtained from the observations of galaxies at
$z\sim2$. The ratio of $\mathrm{EW}(\mathrm{H}\beta)/\mathrm{EW} ([\mathrm{O\,
III}])$ would not be a constant, but depends on various environmental factors,
such as metallicity, stellar mass, and redshift (e.g., \cite{Kewley+2015,
Begley+2025,Laseter+2025}). Thus, the current estimate has a large uncertainty.
}}
With these assumptions, we can infer $\mathrm{EW}(\mathrm{H}\beta) \sim
100$~\AA\ at the peak of log(N/O).  This value is lower than the
theoretical prediction of $\sim 350$~\AA\ from \citet{Inoue2011}, but
both estimates exceed the AGB scenario limit of $\mathrm{EW}(\mathrm{H}\beta) <
40$~\AA\ proposed by \citet{McClymont+2025}. Whether the actual
N-enhanced galaxies exhibit higher $\mathrm{EW}(\mathrm{H}\beta)$ or not remains
unclear, as the samples are currently limited and biased \citep{Zhu+2025}.
Further sample observations are needed to clarify the possible scenarios.

\section{Summary and Discussion} \label{sec:summary}

In this study, we simulated a massive halo formed in a center of a rare density
peak comparable to GN-z11 at $z\sim10$ and the Coma-like cluster at $z=0$. Using
this simulation, we investigated the detailed structure, chemical composition,
and star cluster formation in and around the halo until $z\sim10$.  The features
of the galaxy resemble GN-z11, so we can compare the results with GN-z11 to
understand the origin of the galaxy.

Our simulations are summarized as follows:
\begin{enumerate}
    \item The most massive halo formed at the intersection of large-scale
        filaments and rapid baryon accumulation begins at
        $t\sim420~\mathrm{Myr}; z\sim11$ and continues to
        $t\sim472~\mathrm{Myr};z\sim10$ (figures~\ref{fig:20kpc},
        \ref{fig:2kpc}, and \ref{fig:200pc}).  During this phase, relatively
        low-angular momentum gas accreted onto the center and triggered the
        central starburst.  The density of the central region is sufficiently
        high to allow star formation to proceed before the onset of CCSNe, with
        a local free-fall time of less than $1~\mathrm{Myr}$.
    \item The peak SFR exceeds $100~\Msun~{\mathrm{yr}^{-1}}$ and most of
        star formation occurs in the central region
        (figures~\ref{fig:SFR} and \ref{fig:MassEvolution}). The stellar mass of
        the central region increases more than two orders of magnitude from
        $t=420~\mathrm{Myr};z=11$ to $t=472~\mathrm{Myr};z=10$, resulting in a
        very high star formation efficiency. The SFR and stellar mass of this
        galaxy at $t=444~\mathrm{Myr};z=10.45$ are comparable to GN-z11
        \citep{Tacchella+2023}.
    \item The simulated stellar mass-SFR and the stellar mass-metallicity
        [log(O/H)+12] relations at $t=444~\mathrm{Myr};z=10.45$ show a good
        agreement with GN-z11 \citep{Bunker+2023} (figures~\ref{fig:MstarSFR}
        and \ref{fig:MstarZ}). Also, the galaxy is consistent with the high-$z$
        observations.
    \item The chemical composition of the galaxy center evolves rapidly. During
        the first $10~\mathrm{Myr}$ from the beginning of the central starburst,
        a nitrogen-rich, i.e., high log(N/O), environment is established in the
        gas-phase of the central region $< 10~\mathrm{pc}$ due to stellar winds
        from fast-rotating massive stars. The duration is typically related to
        the lifetime of the most massive stars which result in CCSNe. When CCSNe
        start to occur, the nitrogen-rich environment dilutes and becomes low
        log(N/O) state. The nitrogen-enhanced phase would be highly biased in
        the center and also would be very short-lived.  Rapid gas accretion
        causes the central potential wall to deepen rapidly. Hence effects of
        CCSNe do not immediately halt the central starburst.
    \item Our simulated galaxy contains a number of star clusters. There are two
        populations: ex-situ and in-situ star clusters. Ex-situ star clusters
        formed in the minihalos earlier and accreted to the galaxy later. They
        have higher log(N/O) values and lower metallicities. They are located in
        the outer regions of the galactic halo. On the other hand, star clusters
        of the in-situ origin are located in the inner region having lower
        log(N/O) values and higher metallicities. This is because their natal
        gas is well-enriched in the galactic center. Both are formed on a very
        short timescale ($\sim 10~\mathrm{Myr}$). The star cluster formation
        efficiency reaches $\sim 20\%$ at this epoch.
    \item Some ex-situ star clusters exhibit Na-O anti-correlations. These
        relations are established via rapid pollution by stellar winds in star
        clusters. The anti-correlation is not observed in clusters formed from
        metal-rich gas in our simulation.
\end{enumerate}

\citet{Dekel+2023} proposed a model of feedback-free starbursts operating in
the early Universe.  In their model, high mass galaxies in high-$z$ environments
evolve rapidly with intense accretion of cold, low-metallicity gas, resulting in
highly concentrated low-metallicity gas ($n_\mathrm{th} > 3\times10^3~{\rm cm}
^{-3}$ and $Z < 0.1\Zsun$) at their centers. Since the free-fall time of the
central gas is less than $1~\mathrm{Myr}$, a high-efficiency starburst occurs
before stellar feedback can take effect. This process involves multiple
generations of star cluster formation.

In our simulations, due to the large-scale accretion beginning at $t\sim420~{\rm
Myr};z\sim11$, a very high concentration of low-metallicity gas occurs at the
galaxy center. The central gas density overshoots the critical density of the
feedback-free starburst scenario, $n_\mathrm{th} > 3\times10^3~{\rm cm}^{-3}
$. Consequently, the intense starburst takes place. The analytical model of the
feedback-free starburst predicts the feedback-free starburst phase of
$\sim100~\mathrm{Myr} $ \citep{Li+2024FFB}. Our model exhibits a slightly
shorter duration of the starburst, i.e., $\sim50~\mathrm{Myr}$. This difference
could be attributed to the actual concentration, the adopted mass range for
CCSNe, the limitations of modeling in feedbacks, etc. Also, the galactic scale
star formation in our simulations is monotonic and continuous, although it
consists of multiple small bursts of clumps.

The chemical evolution of our simulated galaxies shows that only a very
early phase of $\sim 10~\mathrm{Myr}$ exhibits a high log(N/O) value due to stellar
winds and this high log(N/O) is diluted by ejecta from CCSNe. If the situation is
similar in high-$z$ environments, nitrogen-enhanced galaxies should be biased
toward younger populations. Indeed, as table \ref{tab:Nenhanced} shows, the stellar
ages are predominantly very young.

It is worth comparing our results with \citet{KobayashiFerrara2024}. They use
one-zone models to describe the chemical evolution of GN-z11. Their results show
that stellar winds from rotating massive stars under a single burst model cannot
reproduce an elevated log(N/O) within a suitable range of $\log(\mathrm{O/H})+12$,
regardless of pre-enrichment. This led them to propose a dual starburst model.
The dual starburst model can reproduce $\log(\mathrm{N/O}) > -0.25$ at suitable
values of $\log(\mathrm{O/H})+12$, while the duration is less than
$1~\mathrm{Myr}$. In contrast, our simulation produces a supersolar log(N/O)
with comparable $\log(\mathrm{O/H})+12$ during a single burst with a
duration of $\sim 10~\mathrm{Myr}$, although the value does not exceed -0.25. A
possible key difference lies in the metallicity of the accreting gas. In our model, the
accreting gas has $\log(\mathrm{O/H})+12 \sim 7.5$ ($\sim0.1\Zsun$) with large variations
(see figure~\ref{fig:phase:Rvel_OH12}). This gas has been preprocessed by
stellar winds, supernovae, and AGBs from Pop III and II/I stars. Because of this
lack of pristine gas, our model balances high log(N/O) with $\sim$ 0.1 $\Zsun$
without requiring dual starbursts.

Nitrogen-enhanced galaxies have high log(N/O) values while maintaining normal
log(C/O) values. Using the yield table of \citet{LimongiChieffi2018}, our
current model produces high log(C/O) values, creating an apparent conflict with
observations. This result is consistent with \citet{BekkiTsujimoto2023}, who used
one-zone models. Similarly, GCs exhibit high log(N/O), but normal log(C/O)
values. This discrepancy can be partially attributed to the
\citet{LimongiChieffi2018} yield table, which several studies have noted
over-production of carbon \citep{Prantzos+2018, Romano+2019, Kobayashi2022,
BekkiTsujimoto2023, Rizzuti+2024}. Future update of the yield table may solve
the issue.

As a final note, we address the current limitations and future prospects. Mass
and spatial resolution are still insufficient to account for the detailed
formation and evolutions of high-$z$ galaxy centers.  Finer mass and spatial
resolution studies while focused on host galaxy/cloud masses much smaller than
ours, show more complicated metal distributions and episodic nature
\citep{FukushimaYajima2024,Sugimura+2024}. Our simulations lack these finer
details.

The treatment of the individual stars can introduce other possible channels such
as very massive stars and tidal disruptions. \citet{Fujii+2024} demonstrates the
possibility of very massive star formation and succeeding intermediate mass
BH formation in the center of a GC by precisely solving the stellar
dynamics. Future high-resolution simulations will be able to directly evaluate
the chemical evolution in the heart of high-$z$ galaxies while considering these
various channels.

\begin{ack}
We thank Xihan Ji for useful discussions. Numerical simulations were carried out
on Cray XC50 and XD2000 at Center for Computational Astrophysics, National
Astronomical Observatory of Japan.
\end{ack}

\section*{Funding}
This study was supported in part by JSPS KAKENHI Grant Numbers JP22KJ0157,
JP21K03614, JP22H01259, JP24KJ0202, JP25H00664, JP25K01046, MEXT as ``Program
for Promoting Researches on the Supercomputer Fugaku'' (Structure and Evolution
of the Universe Unraveled by Fusion of Simulation and AI; Grant Number
JPMXP1020230406).

\section*{Data availability}
The data underlying this article are available upon request to the authors.


\end{document}